**Solute transport predicts scaling of surface reaction rates in porous media: Applications to silicate weathering**

A. G. Hunt[a,b]
T. E. Skinner[a]
Behzad Ghanbarian[b]

[a] Department of Physics, Wright State University, Dayton, OH, 45435, USA
[b] Department of Earth & Environmental Sciences, Wright State University, Dayton, OH, 45435, USA
corresponding author e-mail allen.hunt@wright.edu

**Abstract**

We apply our theory of conservative solute transport, based on concepts from percolation theory, directly and without modification to reactive solute transport. In this approach, chemical reaction rates are limited by solute transport velocities, and these diminish in different ways with increasing length and time scales, both with a cross-over to a steeper slope at larger scales. The results of this theory have already been shown to predict the observed range of dispersivity values for conservative solute transport over ten orders of magnitude of length scale, including two slope cross-over points. We now show that the temporal dependence derived for the solute velocity predicts the time-dependence of the weathering of silicate minerals over twelve orders of magnitude of time scale, while its length dependence agrees with data for the length dependence of reaction rates over five orders of magnitude of length scale. In both cases it is possible to unify lab and field results. In the case of the temporal dependence two slope cross-overs can be identified in elution data; these two slope changes correspond to those seen in the dispersivity. We suggest the possible relevance of our results to landscape evolution of the earth's terrestrial surface.

**Keywords:** Fractal dimension, Percolation theory, Reaction rate, Solute transport, Weathering

**Introduction**

About a decade ago, White and Brantley (2003) published a seminal paper summarizing time-dependent silicate weathering rates. The most important aspect of their results for the purpose of this particular study is that the reaction rates decayed differently at different time scales, making it impossible to predict long-term field reaction rates accurately from shorter term lab experiments. In particular, the rate at which weathering reactions slow steepens significantly with time. In a similar study Maher (2010) showed that there is a related, though distinct, dependence of weathering rates on spatial scales. Furthermore, laboratory results from Hanford studies (Zhong et al., 2005; Liu et al., 2008,2009; Peng et al., 2012; Du et al., 2012) show the same spatial dependence of dissolution rates for Uranium complexes as reported by Maher, and the same temporal dependence as found by White and Brantley.
Understanding weathering rates in natural settings can help us predict soil production and associated denudation rates (Anderson and Anderson, 2010), since slow kinetics of silicate mineral dissolution allows growth of both vascular plants and vital microorganisms (Maher, 2010). In addition, (Maher, 2010) "Slow dissolution of minerals on land and formation of biogenic calcite in the oceans also maintains atmospheric $CO_2$ concentrations and therefore plays an important role in maintaining global temperatures at levels optimal for the presence of liquid water (Berner, 1992). In Earth's past, major changes in rock weathering have coincided with periods of mass extinction (Algeo and Scheckler, 1998; Sheldon, 2006) and reorganization of global biogeochemical cycles (Raymo, 1994; Vance et al., 2009)." Another important implication is to the release of Uranium from Hanford sediments, hitherto poorly understood, that can lead to groundwater contamination (Zhong et al., 2005).

A successful theory for particle surface reactions in porous media, including chemical weathering, may therefore have great relevance. The purpose of this work is to show that application of our existing theoretical description of solute transport (Hunt and Skinner, 2008, 2010a,b; Hunt et al., 2011) to reactive solute transport generates exactly the observed reaction rate dependence for time scales ranging from minutes to millions of years. In addition, it generates the same spatial dependence for reaction rates observed by Maher (2010) and at Hanford, thus providing a unified theoretical description over an enormous range of spatial and temporal scales. The theory is based on the hypothesis that surface reactions are transport limited, giving, in that case, reaction rates that are proportional to solute transport rates. It is known that physical denudation rates are typically proportional to chemical weathering rates (Dixon et al., 2009). With the additional hypotheses that the predicted transport distance is essentially an equilibration distance, and that this distance can be identified as a weathering depth (see e.g., Lin et al., 2012), we can also generate typical landscape denudation rates, and the approximate time scale over which a complete geomorphic reworking of the earth's surface occurs is believed to occur (Anderson and Anderson, 2010).

We emphasize that our theoretical description is not based on the traditional advection-dispersion equation (ADE) (as, e.g., in Lebedeva et al., 2010), nor on the assumption that the process of diffusion is limiting equilibration. The ADE uses continuum concepts to define solute transport as occurring by diffusion (or an analogous process called hydrodynamic dispersion) and by fluid advection. Although the ADE is very frequently used as a basis for modeling solute transport (e.g., Scheidegger, 1959; Bear, 1972), it has a number of defects when its predictions are compared with experiment. These include a failure to generate long-tailed solute arrival time distributions (Berkowitz and Scher, 1995; Cortis and Berkowitz, 2004; Hunt et al., 2011) as well as an inability to produce a dispersivity that is linear in the mean travel distance, as revealed in experiments (Gelhar et al., 1992; Hunt et al., 2011) over 10 orders of magnitude of length scale. These problems have also been pointed out in discussions of the continuous time random walk (CTRW) (e.g., Berkowitz and Scher, 1995). Our analysis of the ADE includes a treatment of the effects of molecular diffusion (Hunt and Skinner, 2010b; Hunt et al., 2011). Nevertheless, it was shown that spatially variable advection, neglecting molecular diffusion, proved sufficient to predict solute dispersion (Hunt and Skinner, 2010a; Hunt et al., 2011; Ghanbarian-Alavijeh et al., 2012), including cases in which the solutes were reactive (Haggerty et al., 2004).

**Theory**

*Background*

Rates of reactions on surfaces of particles in complex porous media can display intriguing dependences on time and length scales. These include power-law decays in reaction rates in both space and time. The fact that the relevant power in the power law depends on whether the reaction rate is represented as a function of distance or of time requires explanation. Further, there is considerable evidence that the power in the temporal decay law changes at larger values of the time scale. Possible causes mentioned in the literature include both evolution of the surface and the possibility that the rates are transport-limited. The latter interpretation is typically based on the assumption of the relevance of diffusion of reaction products (or reagents) through the fluids in the pore space in preventing chemical equilibration. A change in slope with increasing time scale is often assumed to result from changes in specific surface area, either with length or time scales. An analogous assumption regarding a change in architecture of the subsurface with increasing length scales has been made to explain a change in behavior of the dispersivity of conservative solutes as a function of length scale (Neuman and di Federico, 2003; Xu and Eckstein, 1995). This behavior has been recently interpreted (Hunt et al., 2011) as due to a change in the scaling of the solute velocity at a cross-over to universal behavior from percolation theory.

The fundamentals of surface complexation and dissolution in natural porous media are thus not understood in detail, particularly under field conditions. By analyzing a wide range of weathering rate data, White and Brantley (2003) were able to demonstrate a systematic diminution of reaction rates over time scales from days to millions of years. These authors also performed weathering rate experiments over a long enough time span to generate a temporal dependence that was in accord with that deduced from natural processes in situ, *at least for the same time scale*. Specifically, fresh Panola plagioclase weathering rates decayed as the -0.51 power of the time over time scales from about one month to six years. Such a power is essentially indistinguishable from that predicted by diffusion-limited reactions (-0.5) (Doering and Ben-Avraham, 1988). On the other hand, fitting a power-law to the decay of all the silicates reported in all the studies these authors compiled, including times from days to 5 million years, yielded a power of -0.61. As a consequence, extrapolation of the White and Brantley (2003) experimental results on fresh surfaces to time scales of a million years would lead to an overestimation of weathering rates by about a factor 10. The discrepancy originates in a significant slowing of weathering rates (to a slope of -0.7 or greater) that occurred at times exceeding about 1000 years, a result also noted by Maher (2010). Diffusion, as the rate-limiting form of transport, does not provide for a change in the functional form of the temporal dependence of the reaction rate.

If weathering rates are surface-area controlled (Navarre-Sitchler and Brantley, 2007), one may seek explanations for the discrepancies between laboratory and field values of the reaction rates in either scale-dependent surface areas (Navarre-Sitchler and Brantley, 2007), or time-dependent surface characteristics, such as pitting (White and Brantley, 2003). Indeed, high in the soil profile, mineral grain surfaces are pitted (Brantley et al., 1986). Nevertheless, there is evidence that deep in soil profiles, more relevant to the establishment of the weathering front advance rate, the mineral grains are not pitted (Brantley et al., 1986). In our theory, as we show here, the additional slowdown can be attributed to reductions in solute transport velocity with increasing distance of transport rather than changes in surface chemistry.

The relevance of transport to the kinetics of silicate weathering has been discussed for decades (e.g., Muir et al., 1989; Knapp, 1989; Schnoor, 1990; Casey et al., 1993; Kump et al., 2000; White and Brantley, 2003; Li et al., 2008; Raoof and Hassanizadeh, 2010; Maher, 2010; Dentz et al., 2011; Noniel et al., 2012). Already in the review of 2003, White and Brantley considered the possibility that transport limitations could become more important with increasing scale, although they mentioned only diffusive processes. Later investigations appeared to rule out the relevance of transport-rate limited weathering at the scale of a single pore (Li et al., 2008), but experiments ruling out transport limitations at larger scales, which employ stirring techniques (Noniel et al., 2012), seem less convincing because stirring hinders the development of concentration gradients. As several authors have pointed out (Maher, 2010; Raoof and Hassanizadeh, 2010; Dentz et al., 2011), during solute transport, concentration gradients can develop within the pore space, making it incorrect to assume that equilibrium has been achieved throughout. And Maher's (2010) investigation led to the conclusion, "An analysis of weathering rates for granitic material shows that weathering rates depend most strongly on fluid residence times and fluid flow rates, and depend very weakly on material age."

The theory of solute transport must, of course, describe at least the temporal evolution of solute distributions, as we have done (Hunt and Skinner, 2008; Hunt et al., 2011; Ghanbarian et al., 2012). But for the purpose of the present work, we need only discuss the typical solute transport speeds in any detail. The fundamental results of our calculations are in agreement with known results from application of continuous time random walk (CTRW) theory (Scher and Montroll, 1975; Scher et al., 1991). At small enough length scales, where the ADE is valid (Neuman 1990; Hunt et al., 2011), the mean solute velocity is the same as the mean fluid velocity. But at larger length scales, however, the power-law distributions that we obtain are known to introduce several counterintuitive results.

There exist conditions, such as three-dimensional flow through highly heterogeneous materials, where the mean solute travel time cannot be determined as a function of distance–when the parameter $\beta$ in the CTRW is less than 1, this average over all times diverges (Scher and Montroll, 1975; Scher et al., 1991; Sahimi, 2012). By contrast, the measurement of a mean solute transport distance as a function of time *is* possible. Any definition of a mean solute velocity must take this distinction into account. Intuitively, solute transport velocities would seem to be the same as fluid velocities. However, the largest length scale at which solute and fluid velocities are equivalent is probably the scale of a single pore, the largest length scale at which the ADE is valid (Neuman, 1990; Hunt, et al., 2011). At increasing length scales, according to our results, the mean solute velocity diminishes rapidly even when the fluid velocity remains constant. These results for the solute velocity are in strong contrast to predictions of the ADE, though they are consistent with the CTRW (e.g., Scher et al., 1991; Sahimi, 2012), since our values of $\beta$ satisfy the criterion $\beta < 1$. Nevertheless, at all distances, the solute velocity scales with the fluid velocity at the pore scale, the largest length scale at which the ADE is applicable.

*Percolation Theoretic Approach*

We provide only a brief overview of existing theory sufficient for its new application to the present problem. Details can be found in (Hunt and Skinner, 2008; 2010a). These calculations are based on concepts from critical path analysis, an application of percolation theory developed originally to upscale conduction properties of media with wide ranges of local conductance values (Ambegaokar et al., 1971; Pollak, 1972; Hunt, 2001). Note that while the fundamental strategy employed in this theory may be applied to any porous medium, other methods may perform better if the medium is only weakly disordered, such as random close packing of monodisperse glass beads (Hunt and Ewing, 2009).

In the original development (Hunt and Skinner, 2008; Hunt et al., 2011) it was assumed that the pore-size distribution of the medium follows the truncated random fractal model of Rieu and Sposito (1991). The details of the model have only minor influence on the behavior that we investigate here (see the figures), but a concrete model allows us to investigate effects of varying heterogeneity on solute transport and make specific predictions using parameters that can be extracted from experiment. For the Rieu and Sposito model, the probability density function for the pore size distribution is given by (Hunt and Gee, 2002):

$$W(r) = \frac{3-D}{r_{max}^{3-D}} r^{-1-D}, \quad r_{min} < r < r_{max} \qquad (1)$$

where $D$ is the fractal dimensionality of the pore space, $r$ is the pore radius, and $r_{min}$ and $r_{max}$ are the smallest and largest pore radii, respectively. It is important for future reference that use of a power-law pore-size distribution is not the cause of the long tails of the solute arrival time distribution. Note that the largest and smallest pore radii of the distribution correspond also to maximum and minimum hydraulic conductance values, $g_{max}$, and $g_{min}$. If it can be assumed that the pore length is proportional to its radius, as would be required by the assumption of the self-similarity of the pore space, the volume of each pore is the product $r^3$ and a numerical constant that depends on the precise pore geometry (Hunt, 2001). Such a numerical constant can be absorbed into the normalization constant (Hunt, 2001).

The porosity of the medium (proportional to the total pore volume) may be found by integrating $r^3 W(r)$ between limits $r_{min}$ and $r_{max}$ to obtain

$$\phi = 1 - \left(\frac{r_{min}}{r_{max}}\right)^{3-D} \tag{2}$$

in agreement with the result from the original discrete model (Rieu and Sposito, 1991). The theoretical description (detailed in Hunt and Skinner, 2008) uses percolation theory to quantify the frequency of occurrence of potential solute transport paths that span a system of length $x$ with a given rate-limiting conductance $g$ relative to the critical hydraulic conductance $g_c$, valid in the range of minimum and maximum hydraulic conductance values, $g_{min}$ and $g_{max}$. The critical conductance is found by setting the percolation probability equal to the cumulative probability distribution for local conductances greater than or equal to $g_c$. The percolation probability depends on the details of the system (Stauffer and Aharony, 1994), but its primary influence (Hunt, 2001; Hunt and Skinner, 2008) is on the parameter $g_c$, and the principle relevance of $g_c$ is to derivation of the value of a fundamental time scale (Hunt and Skinner, 2008). Note that in a self-similar fractal medium, $g \propto r^3$ (Hunt, 2001; Ghanbarian-Alavijeh and Hunt, 2012). The probability distribution of the rate-limiting hydraulic conductance, $g$, for a given distance from the source, $x$, can be derived in terms of an exponential integral, $Ei$, as (Hunt and Skinner, 2008)

$$W(g|x) \propto Ei\left[\left|1-\left(\frac{g}{g_c}\right)^{\frac{3-D}{3}}\right|^2 \left(\frac{x}{L}\right)^{\frac{2}{\nu}}\right], \quad g_{min} < g < g_{max} \tag{3}$$

where $L$ is a fundamental length scale, which, for weathering rates on particle surfaces, we take to be on the order of a particle size, $g_c$ is the critical hydraulic conductance ($\propto r_c^3$ in which $r_c$ is critical pore radius in the medium), $\nu$ is the correlation length exponent from percolation theory (0.88 in a three dimensional system), and $g_{min}$ and $g_{max}$ are given by (Ghanbarian-Alavijeh et al., 2012):

$$g_{max} = g_c \left[\frac{1}{1-\theta_t}\right]^{3/(3-D)} \tag{4a}$$

$$g_{min} = g_c \left[\frac{1-\phi}{1-\theta_t}\right]^{3/(3-D)} \tag{4b}$$

Here $\theta_t$ is the percolation threshold of the medium, expressed as a volumetric moisture content.

Then, the topology of percolation theory and the distribution of controlling conductances for $g_{min} < g < g_{max}$, was used to calculate the time solutes take to travel across the system on such a path. This time, $t$, has the following form as a function of $g$:

$$t = t_0 \left(\frac{x}{L}\right)^{D_b} h(g) \tag{5}$$

where $t_0$ is a pore crossing time, and

$$h(g) = \frac{D}{3-D} \frac{1}{(1-\theta_t)^{\nu D_b - \nu}} \left[\left(1 + \frac{\theta_t}{1-\theta_t}\right)\left(\frac{g_c}{g}\right)^{1-D/3} - 1\right] \left[\frac{1}{\left(\frac{g}{g_c}\right)^{1-D/3} - 1}\right]^{(D_b - 1)\nu} \tag{6}$$

where $D_b$ is the fractal dimensionality of the percolation backbone. Note that $h(g)$ diverges as a power law for $g = g_c$. <u>The exponents $D_b$ and $\nu$ from percolation theory are critical to the form of the long-tailed distribution and thus the scaling of the solute velocities.</u> Equation (3) and Eq. (5) (Hunt and Skinner, 2008, and subsequent publications) may be combined using $W(t|x)dt = gW(g|x)dg$ to find the arrival time distribution, $W(t|x)$ as a function of position, $x$., where $W(g|x)$ and $dt/dg$ terms are calculated using Eq. (3) and Eqs. (5) and (6), respectively. For this purpose, we invoke a numerical procedure which was described in detail by Hunt and Skinner (2008).

Following Hunt and Skinner (2008) and Hunt et al. (2011), we used the results of Lee et al. (1999) to assign $D_b$ as the exponent for the typical system crossing time. At the length scale of a single pore, where the ADE is valid (Neuman, 1990; Hunt et al., 2011), $t_0$ is proportional to a distance divided by a fluid velocity, making the expression for solute velocities proportional to pore-scale fluid velocities. In three dimensions and for random percolation (appropriate for fully saturated conditions, Sheppard et al., 1999) $D_b = 1.87$. Since $D_b$ is greater than 1, the solute velocity is a decreasing function of both time and of solute transport distance. If $D_b$ were equal to 2, and $h(g)$ could be ignored, the scaling of the solute velocity would be the same as for diffusion. Since these conditions are not fulfilled, we have the result that the solute velocity scaling looks very much like diffusion at first, but changes to a steeper dependence at later times.

When the known statistics of $g$ are applied, Eq. (5) might be thought useful for finding a mean travel time. However, with $D_b = 1.87$, such a mean travel time does not exist, as discussed earlier. When a mean travel time does not exist, one can nevertheless find from an analogous equation to Eq. (5) (Hunt and Skinner, 2008) the mean position, $<x>$ of the solute at any given time from an integral over the derived spatial distribution of the solute. Such a calculation was verified (Hunt et al., 2011) to give a result for the typical (not mean) system crossing time, $t(x)$, in the case of electronic transport in disordered semiconductors and polymers (Bos et al., 1989; Pfister and Griffiths, 1978; Pfister, 1976; Pfister and Scher, 1977; Tiedje, 1984). With such support from an independent verification of this portion of the theoretical output, we apply the same calculations here.

Using $<x(t)>$ one can obtain the mean solute velocity, $v$ ($v = d<x(t)> / dt$). When values from the solute distribution variance are also given, one can then predict all related quantities, such as the longitudinal dispersion coefficient and the dispersivity as functions of either transport distance or of transport time (Hunt and Skinner, 2010a,b; Hunt et al., 2011) and the predictions given in those papers were made with the same calculations that are applied here. The mean solute velocity predicted depends on transport time and can be written,

$$v(t) = \frac{L}{t_0} f\left(\frac{t}{t_0}\right) \approx v_0 \left[\frac{t}{t_0}\right]^{(1-D_b)/D_b} \tag{7}$$

where the function $f$ must be determined numerically via the procedure described below eqn(6) using the probabilistic transformation. Here, $v_0 \equiv L/t_0$. Using the mean solute velocity one can transform Eq. (7) so that $x$ is the independent variable. Note that the approximate equality holds only for certain ranges of time, mostly at smaller time values. The hypothesis here, that reaction rates are solute transport limited, makes them proportional to solute velocities, and provides an unexpected opportunity to verify the theoretical result in Eq. (7). Therefore,

$$R(t) = \frac{R_0}{v_0} \frac{L}{t_0} f\left(\frac{t}{t_0}\right) \approx R_0 \left[\frac{t}{t_0}\right]^{(1-D_b)/D_b} \tag{8}$$

Again, in Eq. (8), the approximate equality holds for specific ranges of time. Also, Eq. (8) can be transformed so that $x$ is the independent variable. After such a transformation, (using Eq. (5)) the approximate power-law result for the reaction rate would be $R_0[x/L]^p$, with $p = 1-D_b$, which can also be written as:

$$R(x) \propto \frac{L^{D_b}}{t_0 D_b} x^{1-D_b} \tag{9}$$

When Eq. (9) is valid (at shorter length scales, as will be seen), reaction rates decay according to the power $1-D_b$.

For particle-sized sources (chemical weathering and uranium dissolution problems), $L$ is a characteristic pore size, as noted above. We typically do not know the value of $t_0$ in Eqs. (8) and (9), although under well-defined laboratory conditions it can be calculated. Thus, the value of $t_0$ provides a relevant scale to the time (horizontal) axis. For comparison with experimental data, it is also necessary to define a vertical scale. This can either be a fundamental rate constant ($R_0$ from Eq. (9)), or a fundamental velocity ($v_0$ from Eq. (8)).

Sometimes one is interested in the dependence of reaction rates on the measurement scale, rather than the distance of transport. Navarre-Sitchler and Brantley (2007), for example, found a power-law increase of apparent reaction rates with the increasing scale of measurement and interpreted this result in terms of a reaction front with a fractal structure. Percolation theoretical treatments also generate such fractal surfaces, and it is, in principle, possible to calculate the geometry of such a reaction front within the same theoretical framework that generates the solute velocities. In particular the perimeter of a percolation cluster in three dimensions has *two* contributions (Kunz and Souillard, 1976; Hunt and Ewing, 2009), one proportional to the square of the radius, and one proportional to the volume of the cluster. The theoretical results from percolation theory apply only when the linear dimension of such a cluster is at least on the order of ten individual units (such as bond lengths, pore separations, or a surface roughness scale, Hunt, 2001) long. Let us define the radius of a large cluster as the correlation length, $\chi$, from percolation theory. The volume of a three-dimensional percolation cluster is proportional to $\chi^{2.5}$ ( $M \propto \chi^{D_M}$ where $D_M = 2.5$ is the universal mass fractal dimension of large clusters near the percolation threshold, Stauffer and Aharony, 1994; Hunt and Ewing, 2009), meaning that the surface area, $A$, has two terms: one is proportional to $\chi^2$ and the second is proportional to $\chi^{2.5}$. Thus we have

$$A \approx C\left(\chi^2 + \frac{\chi^{2.5}}{\chi_0^{0.5}}\right) \quad (10)$$

Eq. (10) contains an unknown numerical factor, $C$, and a scale factor, $\chi_0$. The scale factor represents a length scale above which the fractal properties of the surface of the percolation cluster begin to dominate. This value cannot be smaller than the size of a molecule that probes the surface. In fact, however, the above argument on the minimum cluster size implies the necessity of using a value approximately a factor 10 larger than a molecular diameter, since the cluster surface would not exhibit fractal characteristics at smaller length scales.

**Materials and Methods**

In order to evaluate the proposed percolation theoretical approach, we use different datasets available in the literature. These databases are:

- White and Brantley (2003)
These authors reported weathering rate of silicate minerals, in particular, plagioclase, K-feldspar, hornblende and biotite versus time for both laboratory and field conditions (Tables 4-7 in the original article). The White and Brantley (2003) database covers a wide range of time e.g., $10^{-2}$-$10^7$ years. For the experimental conditions, their flow rate of 10ml/hr by gravity flow through crushed granite corresponds to a pore-scale velocity of 16μm/s, about an order of magnitude higher than assumed for typical weathering conditions in situ.

- Maher (2010)
The data are chemical weathering rates of granitic alluvial materials and some sea-floor sediments as a function of fluid residence time in the range of $10^{-3}$-$10^5$ years, weathering rates versus flow rates as well as surface ages ($10^4$-$10^6$ years) vs. flow rates collected from different published papers (see Maher (2010) and references therein).

- Du et al. (2012)
The sediments collected from the Hanford site was sieved into 4 size fractions e.g., <75 μm, 75-500 μm, 500-2000 μm, and 2000-8000 μm (gravel fraction). A synthetic groundwater with a pH value in the range of 7.8 to 7.9 was applied to study Uranium release from different size fractions of sediment. The data used were concentration as a function of pore volume, which we took as proportional to travel time.

- Liu et al. (2008)
Liu et al. (2008) measured the desorption of U(VI) from contaminated sediments collected at the Hanford 300 area, the Hanford site. In a small column (10.5 cm length by 2.4 cm diameter) sediments with size of 2 mm and less were packed with a porosity of 0.41. In a large column (80 cm length by 15 cm diameter) cobbles and gravels were also included to investigate the effect of such large particles on U(VI) desorption. Using the medium descriptions with given flow conditions, we were able to calculate fluid advection times for single pores, $t_0$, to be 0.08 min. and 0.19 min., for the short and long columns respectively. The columns were saturated from the bottom with synthetic groundwater, and then a constant flow rate was applied to leach the sediments. Particularly with such coarse Hanford sediments, wall flow may not be negligible (Cherrey et al., 2003); investigations (Ghanbarian-Alavijeh et al., 2013) of arrival time distributions (Cherrey et al., 2003) have already shown evidence of 2-dimensional flow attributable to wall flow.

- Liu et al. (2009)
The sediments sampled from the Hanford 300 area, the Hanford site, had been contaminated by U(VI)-containing nuclear fuel fabrication liquid wastes. To study the kinetics of U(VI) desorption, sediment particles less than 2 mm were used. The desorption of U(VI) from sediment was investigated using three U(VI)-free synthetic groundwater (SGW) solutions: SGW2 combination is low pH and carbonate, and high Ca; SGW3 is high in pH and carbonate, and low in Ca; and SGW1 is intermediate between SGW2 and SGW3. In this study, we digitized data of experiments performed under SGW1 and SGW2 solutions. Note that the description of the experimental apparatus does not include flow rates, precluding calculation of $t_0$ in these experiments.

- Zhong et al. (2005)
The sediment was collected from the Oak Ridge site in eastern Tennessee to investigate the microbial immobilization of groundwater U(VI). Uranium(VI) sorption to the sediment and desorption were performed with and without Fe(II) at different pH values. For the purpose of this study, the experiments that showed U(VI) desorption measured under pH = 7 were considered.

-Peng et al. (2012)
Uranium and copper concentrations as functions of depth from the surface of an individual basaltic clast were measured by laser ablation. The depths ranged from 0.6μm to 29μm. Peng et al. (2012) reported that the copper and uranium concentrations were highly correlated, so we plotted them together.

For more information about the datasets used in this study, the interested reader is referred to the original published papers.

**Results and Discussion**

*Summary of relevant prior results*

In previous publications (see e.g., Hunt et al., 2011), we obtained two results for solute distributions that are familiar from applications of the continuous time random walk, CTRW (Montroll and Scher, 1973; Scher and Montroll, 1975; Scher et al., 1991; Berkowitz and Scher, 1995): 1) solute distributions tend to maintain the same shape over time (with an approximately power-law tail) and, 2) the dispersivity is proportional to the distance of transport. The proportionality constant that we obtained was very nearly the value of 0.1 found to summarize field experiments by Gelhar et al. (1992). In a wide range of cases (Berkowitz and Scher, 1995), a spatial dependence of the mean solute velocity different from the mean fluid velocity results, as we find below using several appropriate $D_b$ values. This theoretical construction also delivered the dependence of the envelope of allowed dispersivity values on spatial scales. This dependence was verified in comparison with over 2200 experiments for spatial scales ranging from microns to 100km (Hunt et al., 2011) (Fig. 1 here). It is important that *the same set of predictions* is used for the present comparison (Figs. 2,3, 5 – 14). The experimental data represented in Fig. 1 were taken from the following sources: Baumann et al. (2002), Neuman and Di Federico (2003), Pachepsky et al. (2000), Chao et al. (2000), Kim et al. (2002), Haggerty et al. (2001, 2004), Seaman et al. (2007), Sternberg et al. (1996), Vanderborght and Vereecken (2007) and Danquigny et al. (2004).

We point out that the simplest, most parsimonious, model of heterogeneity possible was applied. In particular, we incorporated only one intrinsic scale of heterogeneity, and the distribution of local hydraulic conductance values (corresponding to pore sizes) was monomodal and of power-law form. The particular distribution chosen (Rieu and Sposito, 1991) can often accurately model water-retention curves (Hunt and Ewing, 2009). One need not focus on the details of the distribution, however, as significant changes in the parameters of the pore-size distribution have only minor influence on the shape of the curves of velocity versus time or distance (see Figures 2,3, and 5, for example).

In such a power-law distribution of pore sizes, one can use the fractal dimensionality, $D$, of the pore space to tune the width of the distribution. A value of $D = 1.5$ will typically correspond to a distribution of local conductances, $g$, varying by as much as an order of magnitude (largest pore sizes roughly a factor two greater than the smallest), while $D =$ 2.95 can produce 5 to 10 orders of magnitude of conductance variation ($g_{max}/g_{min}$) depending on the porosity ($\phi = 1 - \left(g_{min}/g_{max}\right)^{(3-D)/3}$), and $g_{min}$ and $g_{max}$ relate to $r_{min}$ and $r_{max}$. Thus we use "disordered" to refer to a fractal dimensionality of 2.95, and ordered to refer to a fractal dimensionality of 1.5.

To produce Fig. 1, it was necessary to calculate both the first (the mean) and the second moments of the solute distribution (required for the variance). Note that identification of the fundamental length scale, called $L$, as 1m (relevant to solute transport *experiments* for typical initial conditions, Hunt and Skinner, 2010a) was shown to lead to a cross-over in the shape of the dispersivity curve as a function of length scale at around 100m, a value previously chosen (Neuman, 1990; Neuman and Di Federico, 2003), in error (Hunt et al., 2011), as representing a fundamental change in sediment architecture. Note that the role of the experimental set-up in defining the initial conditions and, therefore, the length scale $L$ was already foreseen in work of Rajaram and Gelhar (1991). This cross-over in shape of the dispersivity curve is paralleled by changes in the slope of the velocity with time or distance, as is noted in the later figures. *Because the observed time dependence of weathering rates shows a prominent cross-over to a steeper slope, there would be no possibility of even modeling that time dependence without such slope changes*. But with the smaller distances and velocities involved in reactive solute transport in the vertical direction as well as the smaller scale of $L$ (related to solute sources on the order of particle sizes, rather than a typical experimental apparatus), these cross-overs in field transport occur at much smaller length scales, roughly between decimeters and meters, and, in experiments, perhaps as small as centimeters. Since, in experiments under saturated conditions where flow is three dimensional, solute transport times scale as nearly the distance squared (see below theoretical discussion), an increase in two orders of magnitude of spatial scales is associated with an increase in time scales of nearly four orders of magnitude.

*Current results and comparisons with weathering data summaries*

The hypothesized relevance of solute velocities to transport limited reaction rates triggers the question: How does one obtain a quick estimate for the mean solute velocity as a function of time? In analysis and applications of the theory, the scale factor $\left(x/L\right)^{D_b}$ in Eq. (5) is the dominant input to the velocity, at least at short times. Here, $D_b$ is the fractal dimensionality of the percolation backbone (equal to 1.87 for random percolation in three dimensions, appropriate under saturated conditions, Sheppard et al., 1999). Thus, to lowest order, the mean solute velocity, $v$, should scale with time as $v \propto t^{(1-D_b)/D_b}$. The full solution, involving the derived statistical occurrence of paths with given $g$ (Eq. (3)) and the full time solution (Eqs. (5) and (6)) must be obtained numerically. Accordingly, we represent reaction rates $R$ generally $R = R_0 v(t/t_0)/v_0$ (Eq. (8)), where $v(t/t_0)$ (inferred from Eq. (7)) is the expression for the solute velocity as a function of time. Note that $v_0$ is the value of the solute velocity at the largest scale (typically a pore size) for which the solute and fluid velocities are equal, and $R_0$ is the reaction rate at the same scale. The assumption that chemical equilibration is transport-limited allows the minimum possible value of $t_0$ to be inversely proportional to $R_0$ as well, so that higher values of the fundamental reaction rate can accommodate larger flow velocities under conditions that may nevertheless be regarded as transport-limited.

When $D_b$ = 1.87, the analytical result $v \propto t^{-0.47}$, valid at short time scales, follows. We can express this result in a somewhat more useful form, $v = v_0 (t_0/t)^{0.47}$. In fact, at such short time scales (up to years) our numerical solutions of the velocity as a function of time (shown in Fig. 2a and Fig. 3a) yield $v \propto t^{-p}$; $p = 0.49 \pm 0.02$, which does not differ greatly from the analytical solution. The individual values are shown in bold in Table 1. The value of $p = 0.49$ is also slightly closer than the analytical result of 0.47 to the scaling of reaction rates obtained from experiment on fresh Panola plagioclase (White and Brantley, 2003) (also shown in bold as $R \propto t^{-0.51}$ in Table 1). At longer time scales (Fig. 2b and Fig. 3b) through about a million years, a different result is obtained for the scaling of the velocity, namely $v \propto t^{-q}$, with $q = 0.61 \pm 0.02$ (underlined in Table 1). This exponent is the same as is found for the scaling of the weathering rate (White and Brantley, 2003), $R \propto t^{-0.61}$ (also shown underlined in Table 1). This is an extremely good agreement in a prediction that cannot utilize adjustable parameters. In field observations, roughly double the uncertainty (i.e., 0.04) in the power results, if this uncertainty is obtained by comparison of the various silicates tested. In both results from our numerical calculations, the uncertainty in the exponents derives from the wide range of possible heterogeneity, i.e., conductance (as in permeability) distributions investigated. The widths of these permeability distributions range from about one order of magnitude to up to ten orders of magnitude. Fig. 2 and Fig. 3 show two examples each of the determination of $p$ and $q$.

It should be pointed out that conditions in experiments and field conditions need not always correspond to three-dimensional flow under saturated conditions. Nor is it necessary to assume a fluid flow velocity that is constant over time. We simply take the simplest possible combination of assumptions and investigate the implications of our basic theory. We also point out that two dimensional flow for random percolation has $D_b$ = 1.64, invasion percolation in three dimensions has $D_b$ = 1.46 (appropriate under unsaturated conditions on a drying curve according to Sheppard et al. (1999)), and in two dimensions invasion percolation has $D_b$ = 1.22. These three cases would produce reaction rate decays that follow the power laws, $t^{-0.39}$, $t^{-0.32}$, and $t^{-0.18}$, respectively. They would also produce weathering rind thicknesses that increase with time as $t^{0.61}$, $t^{0.68}$, and $t^{0.82}$, respectively. We mention these additional values here since "many studies of [the thickness of basalt] weathering rinds indicate a $t^{0.8}$ dependence" (Navarre-Sitchler, et al., 2007) rather than the $t^{0.5}$ dependence from diffusion. In any case, the results are summarized for different percolation classes in Table 2. The case of two-dimensional invasion percolation can be relevant for very coarse media when the chief flow is along the walls of the experimental apparatus, but the medium is nevertheless unsaturated (wall flow). Wall flow has been a problem with experiments on some Hanford site sediments (see e.g., Cherrey et al. 2003 and Ghanbarian-Alavijeh et al., 2012). Sahimi (2012) has argued that values of percolation exponents relevant for two-dimensions may be relevant under a wide range of experimental conditions, i.e., when columns are much longer than they are wide. Inferences possible from the present theory and current analysis support only wall flow as a mechanism for two-dimensional transport.

We return to the analysis of Navarre-Sitchler and Brantley (2007) that yielded an *increase* in reaction rate with measurement scale. They suggested that this increase was due to the fractal structure of the reaction front and the corresponding increase in surface area with increasing scale. In their Fig. 3, they find that the reaction rate increases as the 0.33 power of the measurement scale. Navarre-Sitchler and Brantley's interpretation in terms of a fractal surface allowed extraction of a surface fractal dimension $D_s = 2.33$. Consider alternatively the percolation treatment, Eq. (10) with specified fractal dimensionality. The linear dimension of a nitrogen molecule is on the order of $10^{-7}$mm (Navarre-Sitchler and Brantley, 2007); we choose a fundamental length scale an order of magnitude larger, i.e., $10^{-6}$mm. Using $C=10^{-5.5}$, which sets a vertical scale, we then generate the comparison in Fig. 4. This figure also shows the result of Navarre-Sitchler and Brantley (2007). Although their single power may appear to fit the data better, we point out that our comparison involves a single adjustable parameter, i.e., the vertical scale, rather than two such parameters (fractal dimension and numerical prefactor). More generally, if experimental values are known only at three different length scales, a function that is the sum of two such powers may be difficult to distinguish from a single power, particularly in the case of considerable scatter.

*Direct comparison with weathering rate data*

Because comparison of theory with a data summary is only part of the story, we now directly compare the derived velocity, $v = v_0 f(t/t_0)$, as a function of time with the rates of weathering of mineral surfaces, as illustrated in Fig. 5. Note that at large time scales our solute velocity scales with time as a power more nearly equal to - 0.7 than to - 0.6. This is relevant (see Fig. 8 below) for specific Hanford experiments as well as discussion of observations collected by Maher (2010). Maher (2010) interprets the slope at larger time intervals as -1.0, but her data set does not overlap perfectly with that of White and Brantley (2003).

Since our theory does not, as yet, incorporate actual reaction rates or species concentrations, we cannot, in general, give certain units to both the vertical and the horizontal axes. To construct Fig. 5, we took the fundamental time unit in the calculations to be years, used the ratio of the measured $R$ to its maximum value, and adjusted the vertical scale of the calculated $v$ to fall on the experimental curve of the fresh Panola plagioclase, thus interpreting the vertical axis as a solute velocity (with unknown units). We will find that these units are 1 – 10 microns per second. The reaction rates were obtained from White and Brantley's (2003) Tables 4-7 with the exception of FPP (fresh Panola Plagioclase), which were digitized from their Fig. 3. White and Brantley also plotted the FPP sequence simultaneously with the field silicate weathering data with exactly the same results as our Fig. 5.

The effective weathering rates in the field over time scales of several million years are about an order of magnitude lower than a value extrapolated from the fresh Panola plagioclase experiments (FPP). This would correspond to between half an order of magnitude and an order of magnitude shorter length scales than those obtained by extrapolation (relevant for Fig. 8 and its discussion). We used percolation exponents appropriate for three-dimensional flow and random percolation. Note that the choice $D =$ 2.95 appears to be most nearly consistent with the changes in slope of $R(t)$, particularly that from the Costa Rica basalt (Sak et al., 2003).

It can be seen that our predictions track the trends of the data, but with our assumed range of $D$ values ($D = 1.5$ to $D = 2.95$), we do not generate the variability. Therefore, the observed variability in reaction rates is unlikely to be a result of variability in medium physical characteristics or morphology, but is likely due to variability in other factors, such as $R_0$ and $t_0$. Using some results of other authors we consider this issue more closely. Some of this scatter is undoubtedly due to the variability in fluid flow rates, since reaction rates are shown by Maher (2010) (her Figures 4 and 5) to be proportional to flow rates in the transport limited regime (as she asserts, for flow rates greater than 16m/yr = 0.5µ/s). At the other extreme, Molin et al. (2012) assert that transport control is expected only for flow velocities less than 100 µ/s. This slightly larger than two order of magnitude range of flow velocities for which reaction rates are considered to be transport limited appears to introduce a roughly two order of magnitude variability in reaction rates that is due solely to variability in flow rates. Although we do not know the field flow rates, we can calculate the flow rate (and fundamental time scale) associated with the experiments on the FPP by taking the experimental data in White and Brantley (2003) and using that result to estimate error bars on our predictions. In the process we can verify that the fundamental time scale of FPP experiments is roughly in accord with our theoretical predictions as well.

White and Brantley (2003) crushed 750g of granite and placed it in a column of 100cm length and 2.4cm diameter. Using a density of granite of 2.7g/cm$^3$, one finds a porosity of ca. 0.4. At a volume flux of 10ml/hr through this column one can use a mass conservation equation to obtain a pore-scale velocity of about 15µm/s, which, on a logarithmic scale, is near the middle of the range defined by Maher (2010) and Molin (2012). Thus, roughly speaking, one could regard the predicted curve as lying in the center of an allowed envelope that extended up to about an order of magnitude above or below the curve. In Fig. 6 we plot three curves that all correspond to the same scaling function, but with different $v_0$ values that correspond to Maher (2010), White and Brantley (2003), and Molin (2012). This envelope captures most of the data, aside from a few exceptional times. In addition, Navarre-Sitchler and Brantley (2007) assert that weathering rates at a given spatial scale tend to vary by no more than two orders of magnitude when the temperature variability is controlled for. These two orders of magnitude are in accord with our estimated variability due to fluid flow rates. The implication is that the remainder of the variability is due mostly to temperature variations. Thus it is not surprising that one of the times for which inferred weathering rate variability exceeds (by a considerable amount) two orders of magnitude is at 10kyr. In view of the fact that most of the rate values for this time interval were collected at locations that had, 10,000 years ago, been at the margins of the retreating ice sheets (see Tables 4-7 of White and Brantley (2003) and the references therein), a variability that exceeds two orders of magnitude is to be expected.

Given (White and Brantley, 2003) that the particle sizes in the fresh Panola plagioclase experiment range from 0.25mm to 0.85mm, one can estimate a typical pore size as 0.3 times the typical particle size (Gvirtzmann and Roberts, 1991), or about 0.15mm. Thus a typical pore crossing time is about 150μm / 15μm/s = 10s. With a column length of 100cm, one can find between 1000 / 0.85 = 1176 and 1000 / 0.25 = 4000 particles along the length of the core, and a similar number of pores. Core solute *transit* times of between (10s) $(1176)^{1.87}$ = 0.17yr and (10s) $(4000)^{1.87}$ = 1.7yr would be expected, comparable to the range of time values in the experiment (White and Brantley, 2003) that extended from 0.19yr to 6.2yr.

Despite the large scatter in the data, our theory reproduces precisely the results obtained from the statistical analysis of the observed reactions rates, as seen above. Furthermore, now we see that the scatter can be interpreted mostly in terms of temperature and flow effects, both of which would show up in a variability in $R_0$. The magnitude and temporal variability of the scatter around our prediction both make sense in the light of the results of other authors (Maher, 2010; Lin, 2012; Navarre-Sitchler and Brantley, 2007). We therefore assert that our results strongly suggest that the variation in reaction rate with time is due to temperature and fluid flow velocity variability.

Maher (2010) has incorporated data unavailable to White and Brantley in 2003. In her Fig. 2, Maher plots data for the dependence of reaction rates on fluid residence time. The range of time scales is from hours to about 100,000 years. Over this time range, reaction rates reported decline according to the -0.53 power of the residence time. In our Fig. 7, we compare our predictions with the data of Maher (2010). By defining the fundamental time scale in the theoretical calculation as 10 years, rather than 1year, this figure continues the slope of about -0.5 to a time one order of magnitude longer than in Fig. 5. Since the interpretation of *t* in Maher's data sets is not quite the same, being a fluid residence time, rather than a weathering time, it is not clear that the fundamental time scale (set by fluid velocities) need be the same as for the White and Brantley (2003) data, which may provide some justification for this change in the time axis. We have again used $D$ = 2.95 with exponents from three-dimensional random percolation in the comparison, implying that the weathering occurred under saturated conditions.

For an additional test, we refer to Maher's (2010) Figs. 4 and 5, which indicate that for field measurements weathering rates are proportional to fluid flow rates, while flow rates and surface ages are inversely proportional. Further, in Maher's (2010) Fig. 7 it is demonstrated that observed reaction rates decline approximately as the soil thickness raised to the -0.9 power (for thicknesses of roughly 10cm to 1m), as one would expect from using the thickness of the soil cover as a proxy for the transport distance, *x*. Solute velocities are proportional to $x / t \approx x / x^{Db}$ from Eq. (5), giving $x^{-0.87}$. This prediction agrees well with Maher's experimental result (Fig. 7). Note here that our transport distance *x* corresponds to the length scale known as the weathering front advance in papers on weathering rind formation (e.g., Sak et al., 2003; Navarre-Sitchler and Brantley, 2007; Navarre-Sitchler et al., 2007).

We now compare data both from Maher's (2010) Fig. 7 and from Peng et al. (2012), for the length dependence of the reaction rate with theory in our Fig. 8 and find reasonable agreement, again for 3D random percolation and pore-space fractal dimensionality 2.95. However, an accurate comparison of this sort requires putting both results on the same plot (Fig. 8), meaning that it is important to evaluate the scale constants, $x_0$ and $R_0$. We normalized the soil weathering rates to the value quoted in the text of Maher (2010), ($10^{-2.3}$), but we did not have an exact value for $x_0$. Then we used experimental data at laboratory (intraparticle) length scales (Peng et al., 2012). Peng et al. (2012) measured Uranium and Copper concentrations as functions of depth from the surface of an individual basaltic clast using laser ablation. The depths ranged from 0.6μm to 29μm. Peng et al. (2012) reported that the copper and uranium concentrations were highly correlated, so we plotted them together. To this purpose we had to normalize both the copper and the uranium depths and rates to their initial values; those two data series are consequently nearly indistinguishable. Since the requisite value of $x_0$ would refer to intraparticle flow (very short penetration depths), the same $x_0$ could not be appropriate for the data we digitized from Maher (2010). Those data refer to interparticle flow through soils over distances up to a meter with pore sizes roughly a particle radius. Thus the ratio of the two length scales we need is the ratio of a particle diameter to that describing the scale of the flow variability within the particle. We used a ratio $x_0$(interparticle)/$x_0$(intraparticle) = 1mm/1μm = $10^3$, but this argument is only qualitative. If we had used an order of magnitude smaller ratio of fundamental distances, the two data sequences would have followed the same curve; however, we have no reason to prefer that value over the one we did use. The change to a more negative slope of the weathering rate as a function of distance should have the tendency to limit depths of weathering within bedrock, unless physical weathering processes are very effective at removing the surface and preventing chemical processes (solute transport) from limiting the weathering.

*Weathering summary*

In the following discussion, we apply the hypothesis relating solute transport distances to weathering depths. For smaller length scales, the solute velocity (as shown in Figs. 2a and 3a and Table 1) could be determined fairly accurately by the analytical scaling result, $v(t) \propto (t/t_0)^{(1-D_b)/D_b}$. This means that the solute transport length scale, $x \approx t^{1/D_b}$, (with $D_b$ = 1.87 for 3D flow) can be determined rather quickly for various time scales given by experiment.

Consider a saturated hydraulic conductivity value of $10^{-4}$cm/s (the median order of magnitude for crustal materials, Anderson and Anderson, 2010), and gravity flow. This combination would be consistent with a single pore flow rate of about 1 μm/s, within the fluid velocity limits suggested by Maher (2010) and Molin (2012) as bounding the transport-limited regime. According to Eq. (7) an identical solute transport rate at the scale of a single pore of this size results.

Using our results for the solute velocity, a relevant solute transport distance (compare again with weathering front advance) of 1μm at 1s would translate at 1 yr to a relevant distance of about a half a centimeter (roughly an order of magnitude smaller than solute velocities in the fresh Panola plagioclase experiments, White and Brantley (2003)), White and Brantley (2003)), at 100kyr to a spatial scale of about 4m (compare Maher's (2010) model results yielding 250 kyr for 2m), and at 1.5 million years to approximately 20m. However the nearly one order of magnitude decrease in effective solute velocity, shown in Fig. 5 as occurring at time scales between about 1kyr and 1Myr, would typically reduce this weathered depth to several meters at 1.5Myr, and to about half a meter at 100kyear. Not only would this latter value be consistent with the Maher (2010) model results, it would also improve the correspondence with the results of Lin et al. (2012), in which total solute transport distances of up to about a decimeter are associated with time scales of around 100kyear. By the commonly observed equivalence between chemical weathering depths and physical denudation thicknesses (Dixon et al., 2009), weathering rates in meters per million years lead to similar denudation rates for weathering limited conditions, as obtained through cosmogenic and other methods of dating (Anderson and Anderson, 2010). Accordingly, for weathering rate limited conditions, grain size transport distances at 1second (where equilibrium between mineral surfaces and the fluid can be assumed) can be linked to common landscape surface losses on the order of meters to tens of meters over a time scale roughly equal to the geomorphic age, 5Myr, of the earth's surface (Anderson and Anderson, 2010).

*Direct comparison with reactive solute transport experiments*

Now consider the dependence of Uranium reaction rates on time scales from various Hanford experiments. We start with data from Liu et al. (2008) because these authors give sufficient information for us to calculate $t_0$, even if we cannot determine what the initial reaction rate should be. Their data were collected from both a short and a long column, but the medium characteristics were the same for both, and the sediments involved were very coarse. The flow rates differed substantially, however. The process for calculating $t_0$ involves conversion of a column fluid velocity to a pore-scale fluid velocity, $v_f$, and then finding the ratio of a typical pore size to $v_f$. We found a geometric mean particle size (excluding the gravel fraction of the soil) and assumed that the associated pore size was a factor 0.3 smaller (Gvirtzman and Roberts, 1991). These methods are trivial and summarized elsewhere. In any case, we could generate absolute time scales for both the short column and the long column experiments that are shown in Figs. 9 and 10 and establish that neither experiment was continued to times long enough to reach the slope change. In order to plot all breakthrough curves for the same experimental apparatus on the same graph, we took the ratio of the concentration (assumed proportional to the reaction rate $R$) to its largest value and $t$ to its smallest value in each experiment. Then the logarithm of the ratio of the earliest arrival time to the calculated $t_0$ defines the onset of the data. The values of $t_0$ were 0.08min and 0.19min, for the short and long column experiments, respectively. To generate the exact correspondences we had to adjust a single constant in each graph, the value of $R_0$. We used again $D = 2.95$. Notice that $t_0$ for the short column is 4.8s, approximately half the fundamental time scale obtained for the fresh Panola plagioclase experiment of White and Brantley (2003).

The short column from Liu et al. (2008) (Fig. 9) is consistent with the predictions from two-dimensional invasion percolation, indicating that the relevant portion of the medium is not saturated. In very coarse Hanford sediments it is known that flow along the wall of a sediment core can dominate due to the particularly large pore spaces formed along the boundary between the sediment and the wall (Cherrey et al., 2003). In another case (Ghanbarian-Alavijeh et al., 2012), we found that it was possible to predict the entire solute arrival time distribution (Cherrey et al., 2003 ) using two-dimensional values of relevant percolation exponents and system parameters extracted from pressure-saturation curves. The interpretation is based on the fact that for coarse soils very large pores can be found at the wall of the core. It may be difficult to force 100% saturation of these pores; if they form only 5% of the pore space, even at 80% saturation the system saturation will be 99%., and difficult to distinguish from 100%.

The long column from Liu et al. (2009) (Fig. 10) is consistent with predictions from three-dimensional random percolation, indicating that the entire medium was saturated.

Fig. 11 shows software generated fits to the data in Figures 9 and 10 and demonstrates that the slope values extracted, 0.17 and 0.46 are almost identical to those in the Table 2 for 2D invasion, 0.18, and 3D random percolation, 0.47, respectively.

The most important aspect about the calculations of the absolute time scales is that they demonstrate that the same functional form can apply to experiments with greatly differing time values.

An interesting dependence of Uranium(VI) desorption concentrations on time was obtained by Liu et al. (2009). We digitized the data from Figs. 1-2 of that work. In order to make a comparison with their data, it is necessary to assume that the observed concentrations are proportional to the solute velocity, just as we assumed above that the inferred reaction rates represented a solute velocity. The time span was much shorter than the data summarized by White and Brantley (2003), however, covering only minutes to hours. This data sequence appears to show (Fig. 12) the same kind of cross-over to a steeper slope as seen in White and Brantley's (2003) data. This is generally consistent with the value of the controlling rate constant, $10^{-4.29}$, taken from Table 1 of Liu et al. (2009), which is five to six orders of magnitude greater than the largest lab values reported by White and Brantley (2003). Consequently, experimental conditions with much greater fluid velocities would still allow for transport-limited reaction rates. If we accordingly change the time axis scale from Fig. 5 by a factor of about $10^5$, and plot the observed Uranium concentrations as a function of time we produce Fig. 12. Reference to Fig. 5 shows that the individual weathering rate results cover only the initial and final slopes, not showing the actual behavior at the predicted slope cross-over. But the Uranium experiments (Liu et al., 2009) shown in Fig. 12 appear to show the slope cross-over. We again used 3D random percolation exponents and the same medium model, $D = 2.95$.

Next, consider the data for Uranium elution from Du et al. (2012). These authors report experiments on Uranium dissolution that show a dependence of Uranium concentration that diminishes according to a power of the transport time. The data were collected from column outflow for five different media with differing particle sizes. The data are plotted in Fig. 13 and compared with theory. Each theory line uses the solute velocity as calculated for $D = 2.95$, with 3D flow under saturated conditions (random percolation). The difference in the three predictions lies solely in the choice of the $t_0$ and $R_0$ values. The comparison demonstrates that the three columns with smaller grain sizes can be predicted well using our calculation. For the largest grain sizes (2 to 8 mm) the small number of potential solute paths accentuates fluctuations and renders predictions based on mean solute velocities less useful. Note the systematic trend to smaller values of $t_0$ with increasing particle size, as might be expected for intergranular transport (larger pore sizes, more rapid flow).

Owing to the fact that the interparticle fluids were stirred, Du et al. (2012) interpreted their scale-dependence in terms of diffusion-limited transport within the particles. This interpretation is in accord with the understanding of Noniel et al. (2012), whose stirred experiments did not show evidence of transport limited behavior at larger length scales. Consequently, these authors interpreted the results in terms of intra-particle diffusion near the percolation threshold. Diffusion near the percolation threshold (Du et al., 2012) also leads to reasonable agreement with experiment. Nevertheless, we argue that time dependence of the solute transport over lengths equivalent to thousands of particle separations (as in the case of the smallest particles) is not reflecting the dynamics of the transport within the particles, but rather, the tortuosity of the paths through the porous medium as a whole. In fact, if diffusion within the individual particles were the only potential cause of an anomalous transport coefficient, an effective-medium treatment using the limiting value (at the particle radius) of the diffusion constants for the particles and a different diffusion constant for the medium between the particles would be appropriate. Consider also the data for the medium composed of the largest particles. In this one case the experimental length scales do not significantly exceed the diameter of the largest particle, meaning that intraparticle processes can dominate. But here the power observed is nearly -0.5, typical of three-dimensional advective solute velocities at the shortest time scales. These reasons lead us to favor our interpretation. We again used 3D random percolation with the same medium characteristics, $D = 2.95$.

We consider the comparison with the results of Du et al. (2012) to be potentially the most important of this study. In the comparison with Liu et al. (2009) it was found possible to generate one possible slope change with the appropriate choice of the scale factors on both axes, i.e., $t_0$ and $R_0$. But in the comparison with Du et al. (2012), two slope changes are generated by the choice of these two scale factors; thus not only are all three slope values reproduced, but the appropriate ratio of time scales for the two slope changes is as well. Thus four potentially independent quantities are developed from only two parameters.

Zhong et al. (2005) obtain a remobilization rate that decays as $t^{-0.53}$ (with $R^2 = 0.89$) or $t^{-0.42}$ (with $R^2 = 0.61$). Such a dependence on time would be in accord with either diffusion as a limiting mechanism, or with our results for transport-limited reaction rates. We compare their remobilization rate with our solute velocity predictions in Fig. 14 and find reasonably good agreement. We again used 3D random percolation with the same medium characteristics, i.e., $D = 2.95$.

*Discussion*

We consider first the potential relevance of our calculations based on transport limitations to weathering in extreme environments. We quote Maher (2010): "Systems with surprising degrees of chemical weathering include deep-sea sediments (Maher et al., 2004, 2006b; Wallmann et al., 2008), saprolites (White et al., 2001; Price et al., 2005) and aquifers (Zhu, 2005). The alternative driving forces in these systems, which may be partly to entirely dominated by diffusive transport, are: 1) the precipitation of secondary minerals (Maher et al., 2006b; Maher et al., 2009; White et al., 2009); and 2) the capacity of microbial activity to sustain departures from equilibrium via reaction networks (Aloisi et al., 2004; Maher et al., 2006b; Wallmann et al., 2008)." Note that data from both Maher 2006ab and Wallmann et al. (2008) are included in the analysis and do not appear to deviate from our predictions. Thus our theoretical formulation appears to account for weathering in extreme environments.

"Transport limitation appears to begin as fluid residence times exceed approximately 2 days and flow rates exceed 16 m/yr." (Maher, 2010) The latter corresponds to flows of 0.5μm/s, which is half the value we chose for Fig. 5. If we generously allow flow rates down to about 3m/year, then all the data that are found above a curve one order of magnitude below the theoretical curve in Fig. 5 could be transport limited. From Maher's (2010) abstract "Over moderate fluid residence times from 5 days to 10 yr, characteristic of soils and some aquifers, transport-controlled weathering explains the orders of magnitude variation in weathering rates to a better extent than material age." (Maher, 2010) This time period includes only the first set of data points in Fig. 5 but makes clear that Maher's inferences based on assumption of diffusive control (with a continuation of the slope predicted at short times), are not relevant if our transport theory is correct. In fact, we can infer that transport control is relevant at all times in the picture, meaning that it is also relevant for experiments in all media. Finally, again from Maher's (2010) abstract, "For fluid residence times greater than 10 yr, characteristic of some aquifers, saprolites, and most marine sediments, a purely thermodynamic-control on chemical weathering rates sustains chemical weathering." This conclusion is based partly on her Fig. 2A, which we reproduce here as Fig. 7. Note that there is no slope break in the data (though there would be in Maher's predictions) at any time period on this graph and that all the data included are consistent with our prediction. Also important is that these data include seafloor weathering results from Wallmann et al., 2008, and Maher et al., 2006a.

Note that Maher (2010) ascribes significance to her analyses showing nearly the same reaction rate dependences on both time and distance, at least in what is there denoted as the transport-limited regime. If these two dependences were the same, this would imply that solute velocities were scale-independent. But if these velocities are scale independent, then there is no reason for transport-limitations on reaction rates to introduce the observed scale dependence. In the case, however, that our analysis is correct, we should expect different powers in these two functions, since the solute velocity is not independent of length scale, even though at small enough length scales it does reduce to the fluid velocity. In our view, analysis of the temporal (Figs. 2,3, 5-7, and 9-14) and spatial dependences (Fig. 8) of the reaction rates show that the two different representations of the data exhibit distinct functional dependences, in accord with our theory.

It should be reiterated that the slope cross-overs in the velocity (see Figs. 4-14) also generate slope changes in the dispersivity (Fig. 1), since the dispersivity is the ratio of the longitudinal dispersion coefficient to the velocity. While the precise cause of these changes is not understood at this time, the changes in the solute velocity at the time scales seen could have been anticipated by analysis of the dispersivity. Also, we suggest that it is of great importance that results of individual experiments in both the dispersivity and in reaction rates track the corresponding changes in slope that are predicted.

**Conclusions**

We tested a potential common theoretical description for a large number of different observations regarding chemical reaction rates in porous media. The results include both length and time scale dependences of experimental reaction rates as well as length and time dependences of weathering rates observed in the field. The connection is based on a single theoretical result for the mean solute velocity as derived in a percolation-based analysis. The result of this theory is that characteristic solute velocities typically diminish at length scales beyond a pore scale. Two simple hypotheses, that chemical weathering rates are mostly transport-limited, and that the transport distances control the depth of the surface weathered layer through the advance of the weathering front, suffice, within this new theoretical description of solute transport, to predict observed time and length dependences of weathering rates in the earth's crust and are generally compatible with the thickness of the surface weathered layer. The hypotheses are in accord with the known results that solute transport at the pore scale is described by the advection-dispersion equation (Neuman and Di Federico, 2003), but that at larger scales heterogeneity controls transport through spatially variable advection (Hunt et al., 2011). They are also generally compatible with the interpretation of the origin of the surface weathered layer and its time scale for production and removal (Dixon et al., 2009) under conditions of weathering rate control. Our theory of solute transport is well established for conservative solute transport (Hunt and Skinner, 2008, 2010a,b; Hunt et al., 2011, Ghanbarian-Alavijeh et al., 2012), but this current study is the first investigation of its potential relevance for reactive solute transport.

Specific tests performed here compared results of the temporal and spatial dependences of the solute velocity, considered a proxy for reaction rates, with data from White and Brantley (2003), Maher (2010), and Sak et al., (2003) for chemical weathering rates, and with data for the temporal and spatial dependence of uranium dissolution, deposition, and transport in Hanford sediments from Zhong et al. (2005), Liu et al. (2008, 2009), Peng et al. (2012), and Du et al. (2012). <u>Our comparison between theory and experiment used a pre-existing calculation available in the literature, with the functional dependence already published</u>. For each data set analyzed, we applied the same medium characteristics, i.e., the same porosity, the same critical volume fraction for percolation, and the same pore size distribution, as well as the same assumption of the relevance of exponents from 3-D random percolation (three-dimensional saturated flow), *indeed the identical theoretical calculation and resulting curve*. With this single curve we could generate accurately the experimental results of every one of these authors! We needed only to use different values of relevant time and spatial scales, some of which we were able to calculate from information provided. The only exception to the above uniformity was the short column of Liu et al., 2008, which exhibited evidence that the dominant transport was by wall flow under unsaturated conditions, instead of bulk three-dimensional flow under saturated conditions. The implication of wall flow was reported by Ghanbarian et al. (2012) when data from the full arrival time distribution of a conservative solute (Cherrey et al., 2003) was analyzed. Even though different flow conditions were inferred from the short and long columns, we could use the same medium characteristics for both experiments.

Known results for the dependence of cluster surface area on cluster radius from percolation theory generated, using one adjustable parameter, the dependence of weathering rates (due to changing surface area) on measurement scale (Navarre-Sitchler and Brantley, 2007).

The results of the study suggest that a great deal of the perceived complexity of a very complicated problem in reactive solute transport can be related to known behavior of non-reactive solute transport. That behavior has been understood only recently (above references for conservative solute transport), but had already been modeled using the continuous time random walk (CTRW) (Scher and Montroll, 1975; Scher et al., 1991; Scher and Berkowitz, 1995). Our results are verified throughout, although with the caveat that final confirmation awaits clarification of the actual values of fluid velocities. Thus, our most important immediate need is to develop means, applicable to any particular observation, to assign the appropriate time scale for solute transport at the pore scale, thereby generating a scaling velocity, and making true predictions of chemical reaction rates possible. While such methods are already available in the case of many lab experiments (and were indeed calculated when sufficient information was available), a solution is not as obvious for field observations of weathering rates.

**Acknowledgement**
This research was supported by the US Department of Energy (DOE) Biological and Environmental Research (BER) through the Subsurface Biogeochemical Research (SBR) Science Focus Area (SFA) program at Pacific Northwest National laboratory (PNNL). We acknowledge the Pacific Northwest National Lab (PNNL) financial support from Battelle contract 154808.

Table 1. Comparison of scaling of solute velocities with reaction rates.

| Experiment[*] | | | Theory | | |
|---|---|---|---|---|---|
| Mineral | Time-scale (years) | Slope | D | Early Slope | Full Slope |
| Fresh Panola Plagioclase | $6 \times 10^0$ | **-0.51** | 1.50 | **-0.52** | -0.61 |
| Plagioclase | $3 \times 10^6$ | -0.566 | 2.50 | **-0.49** | -0.58 |
| K-Feldspar | $3 \times 10^6$ | -0.647 | 2.90 | **-0.48** | -0.62 |
| Hornblende | $3 \times 10^6$ | -0.623 | 2.95 | **-0.47** | -0.63 |
| Biotite | $5 \times 10^5$ | -0.603 | | | |
| Average | | | | -0.49 | -0.61 |

[*] Experiments are from White and Brantley (2003).
[**] The average of underlined values.

Table 2. Analytical results related to the scaling of solute transport time with distance.

| Percolation class | $D_b$ | $v(x)$ | $v(t)$ | $t(x)$ | $x(t)$ |
|---|---|---|---|---|---|
| 2D Random | 1.64 | $x^{-0.64}$ | $t^{-0.39}$ | $x^{1.64}$ | $t^{0.61}$ |
| 3D Random | 1.87 | $x^{-0.87}$ | $t^{-0.47}$ | $x^{1.87}$ | $t^{0.53}$ |
| 2D Invasion | 1.22 | $x^{-0.22}$ | $t^{-0.18}$ | $x^{1.22}$ | $t^{0.82}$ |
| 3D Invasion | 1.46 | $x^{-0.46}$ | $t^{-0.32}$ | $x^{1.46}$ | $t^{0.68}$ |

**Figure Captions**

Fig. 1. Comparison of dispersivity values of over 2200 experiments from Pachepsky et al. (2000), Danquigny et al. (2004), Seaman et al. (2007), Kim et al. (2002), Sternberg et al. (1996), Chao et al. (2000), Baumann et al. (2002), Haggerty et al. (2002, 2004), Vanderborght and Vereecken (2007), for the same variability in heterogeneity as in Figs. 2-5 below. The number in parentheses is the number of experiments. Note that the use of the 1m fundamental spatial scale is not supported on the micromodel experiments (Baumann et al., 2002), for which we chose a length scale of ca. 200μm. Such a revision is equivalent to sliding the entire curve down the diagonal three and a half orders of magnitude. Inv stands for invasion percolation choices; all others are random percolation.

Fig. 2. (a) Determination of the exponent of the power law for the solute velocity as a function of time over the first five orders of magnitude of time scale, and (b) Determination of the exponent over the entire ten orders of magnitude of time scale. Media with wide ranges of local hydraulic conductances are chosen ($D = 2.95$).

Fig. 3. (a) Determination of the exponent of the power law for the solute velocity as a function of time over the first five orders of magnitude of time scale, and (b) Determination of the exponent over the entire ten orders of magnitude of time scale. Media with narrow ranges of local hydraulic conductances are chosen ($D = 1.5$).

Fig. 4. The measurement scale dependence of reaction rates. After Navarre-Sitchler and Brantley (2007). Watersheds are plotted as open triangles, soil profile scale as open squares, and weathering rinds as open diamonds. BET surface area-normalized laboratory rates are plotted as closed circles. The straight line is the interpretation of Navarre-Sitchler and Brantley (2007) in terms of a surface fractal dimensionality $D_s = 2.33$, while the curve is the percolation theoretical result (see text under *Current results and comparisons with weathering data summaries*).

Fig. 5. Comparison of the numerically obtained solute velocity as a function of time over 9 orders of magnitude of time scales with weathering rates over the same range of time scales. Data for weathering rates compiled by White and Brantley (2003). FPP stands for Fresh Panola Plagioclase, a reference to laboratory experiments conducted over six years. In order to make this comparison, we made the identification of $t = 1$ in our time scale with one year, and multiplied reaction rates by a factor of approximately $10^{12}$ to obtain velocities (similar in magnitude to White and Brantley's (2003) normalization). The velocities obtained are represented in terms of arbitrary units, but if the fundamental length and time scales suggested in the text are applied, the velocity at the pore scale would be about 1μ/s and the solute velocity after 1 year would be about 5mm/year. Continuing the velocity scaling to larger time scales leads to a solute velocity of approximately $5 \times 10^{-5}$ mm/year after 50 kyear. Data for weathering rind development from Lin et al. (2012) referred to as volcanic in the figure, yield rates of rind thickening at approximately this value at about 50kyear, allowing us to place these data in the figure directly, without adjusting either the horizontal or the vertical scales, as long as the velocity scale is interpreted as mm/year. We leave the vertical scale as an arbitrary

velocity, however, since we cannot verify solute velocities in field weathering data (White and Brantley, 2003).

Fig. 6. Same as previous figure except that a reasonable uncertainty in reaction rates due to acceptable variability in fluid velocity is included. Also, only the more heterogeneous solution ($D = 2.95$) is included as a prediction. Lower (upper) bound on fluid velocities compatible with transport limited reaction rates is given in Maher (2010) (Lin et al., (2012)).

Fig. 7. Comparison of theoretical solute velocity as a function of time with weathering rates as a *function of fluid residence times* (data digitized from Maher, 2010, Fig. 2a). Note that, for this series of experiments, it was necessary to shift the theoretical curve more than one order of magnitude to the right as compared with Figure 5 (consistent with a correspondingly greater fluid velocity); otherwise the predicted slope break would not occur at time scales larger than those observed. We included Maher's (2010) designations; transport-limited for residence times up to a year, and saturation control at larger time scales.

Fig. 8. Comparison with experiments and field data of the theoretical reaction rate as a function of length scale over 6 orders of magnitude of distance (four orders of magnitude of scaled distance, $x/x_0$). Experimental data are from laser ablation studies of the depth dependence of Uranium and Copper concentrations from a single particle (Peng et al., 2012). Field values digitized from Fig. 7 in Maher, 2010. The experimental reaction rate is normalized to the initial U and Cu concentrations (at or near the surface), but for the field data (Maher, 2010), to the given value of $10^{-2.3}$. Experimental and field data are presented together using distinct values of $x_0$: $x_0 = 1\mu m$ for intraparticle data but 0.1mm for the interparticle data, a ratio of fundamental transport distances of only 100. Had we taken the ratio to be 1/1000, all data would fall on the same curve.

Fig. 9. Comparison of theoretical solute velocity as a function of time with Uranium desorption data from the short column described by Liu et al., 2008. The various breakthrough curves were all normalized according to their initial times and concentrations. This set of data is the only one analyzed here that is described by 2-D invasion percolation exponents, implying that in this case flow was concentrated along the walls of the column under unsaturated conditions.

Fig. 10. Comparison of theoretical solute velocity as a function of time with Uranium desorption data from the long column described by Liu et al., 2008. The various breakthrough curves were all normalized according to their initial times and concentrations. Here the usual random percolation exponents for 3-D flow were applied.

Fig. 11. The same data are presented as in Fig. 8 and Fig. 9, but this time the exponents, 0.17 and 0.46, were determined by commercial software (Excel). These values are essentially equal to the entries in Table 2 for temporal scaling of the velocity, 0.18 and 0.47, for 2-D invasion and 3-D random percolation, respectively.

Fig. 12. Comparison of the theoretical solute velocity as a function of time with Uranium desorption data (Liu et al., 2009). The procedure for the experiment is described in detail in Liu et al., 2009; the various breakthrough curves, however, correspond to renewed elution after hiatuses from stopped flow. We note that the displayed agreement between theoretical and experimental slope changes is possible only if the time scale chosen for theory is correct.

Fig. 13. Comparison of the theoretical solute velocity as a function of time with elution data of Du et al. 2012. The different data sequences are for media with different particle sizes. As expected, the fundamental time constant diminishes with the coarsening of the particles, commensurate with larger pores and more rapid flow. We again used 3-D random percolation exponents, consistent with three-dimensional flow and saturated conditions, as well as the choice $D = 2.95$. The fundamental time scale and reaction rate were adjusted for the best fit.

Fig. 14. Comparison of the theoretical solute velocity as a function of time with iron concentrations from Fig. 1b of Zhong et al. (2005). The other figures were omitted since, in most cases, the iron concentrations did not exhibit a systematic behavior, but, rather fluctuated around a single value, or increased steadily with time. In the latter case, it would be theoretically possible to investigate the difference between the iron concentrations at a given time and their final values (which steadily diminishes), but the final values cannot easily be inferred.

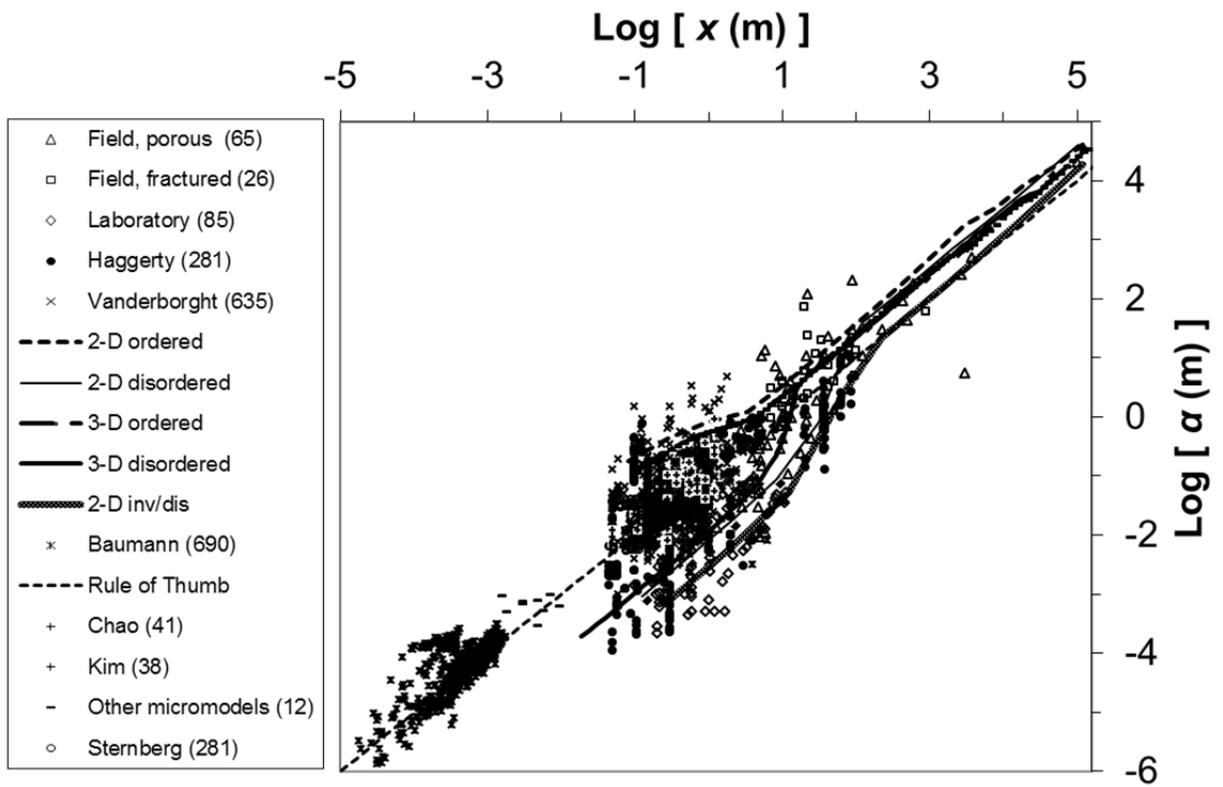

Fig. 1

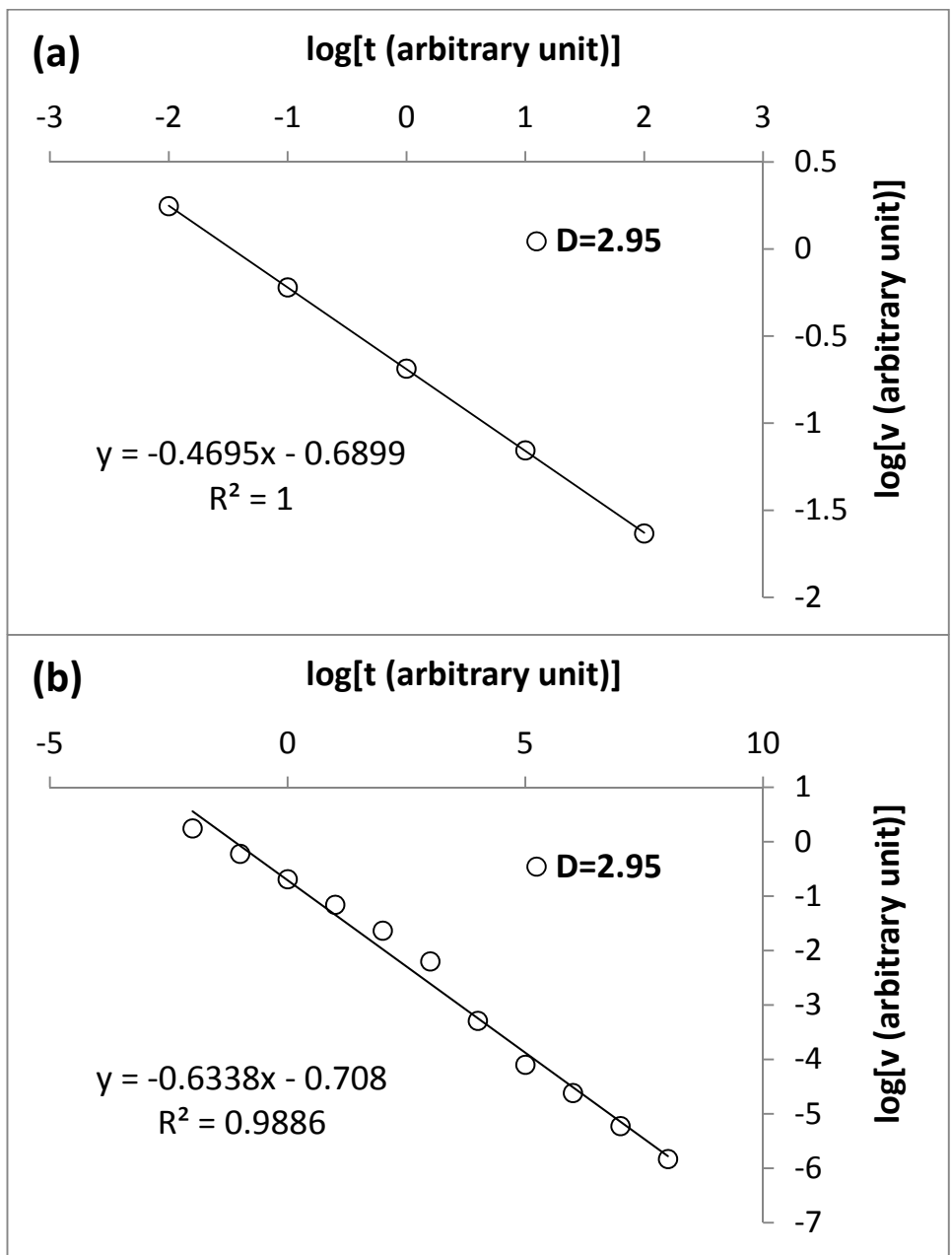

Fig. 2

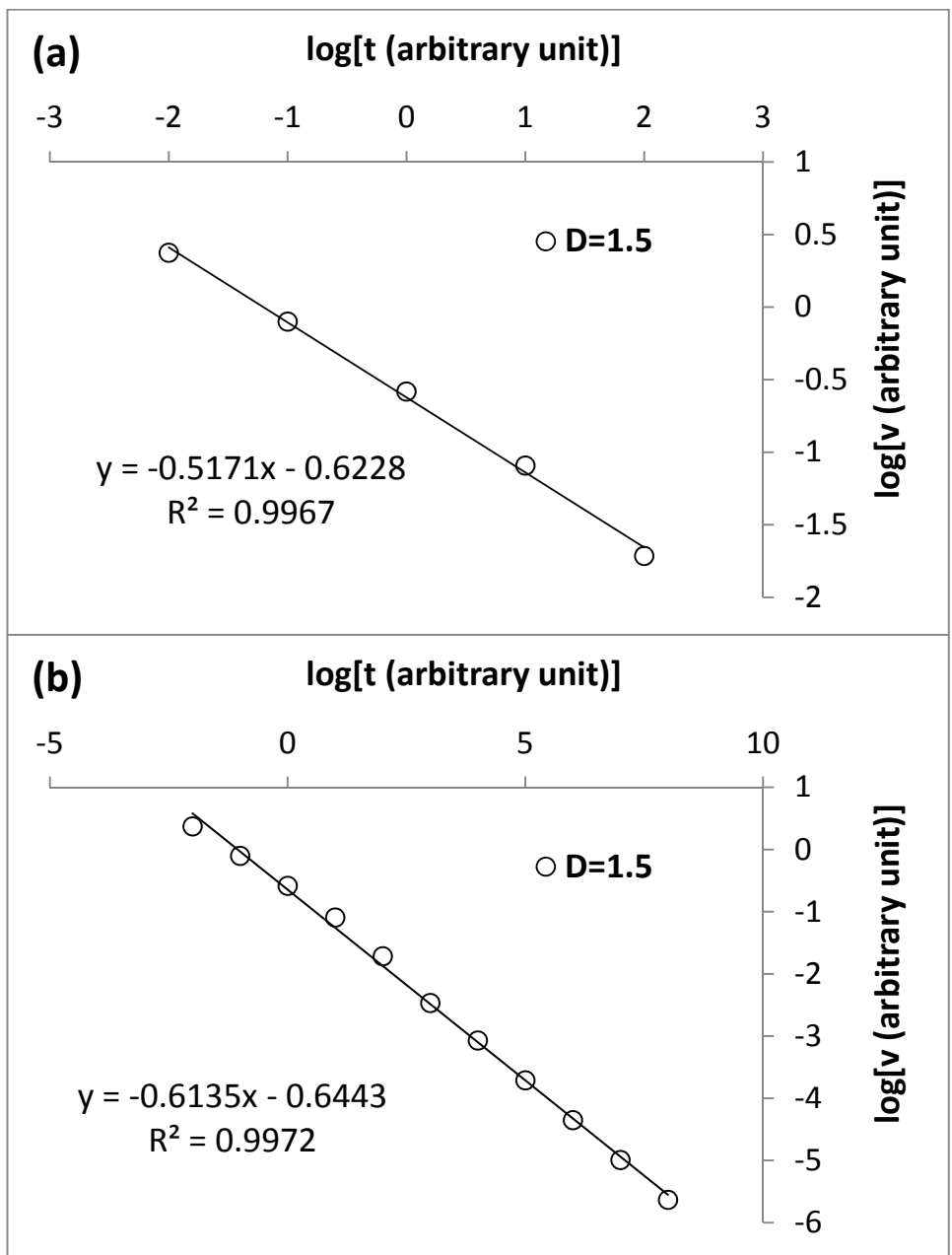

Fig. 3

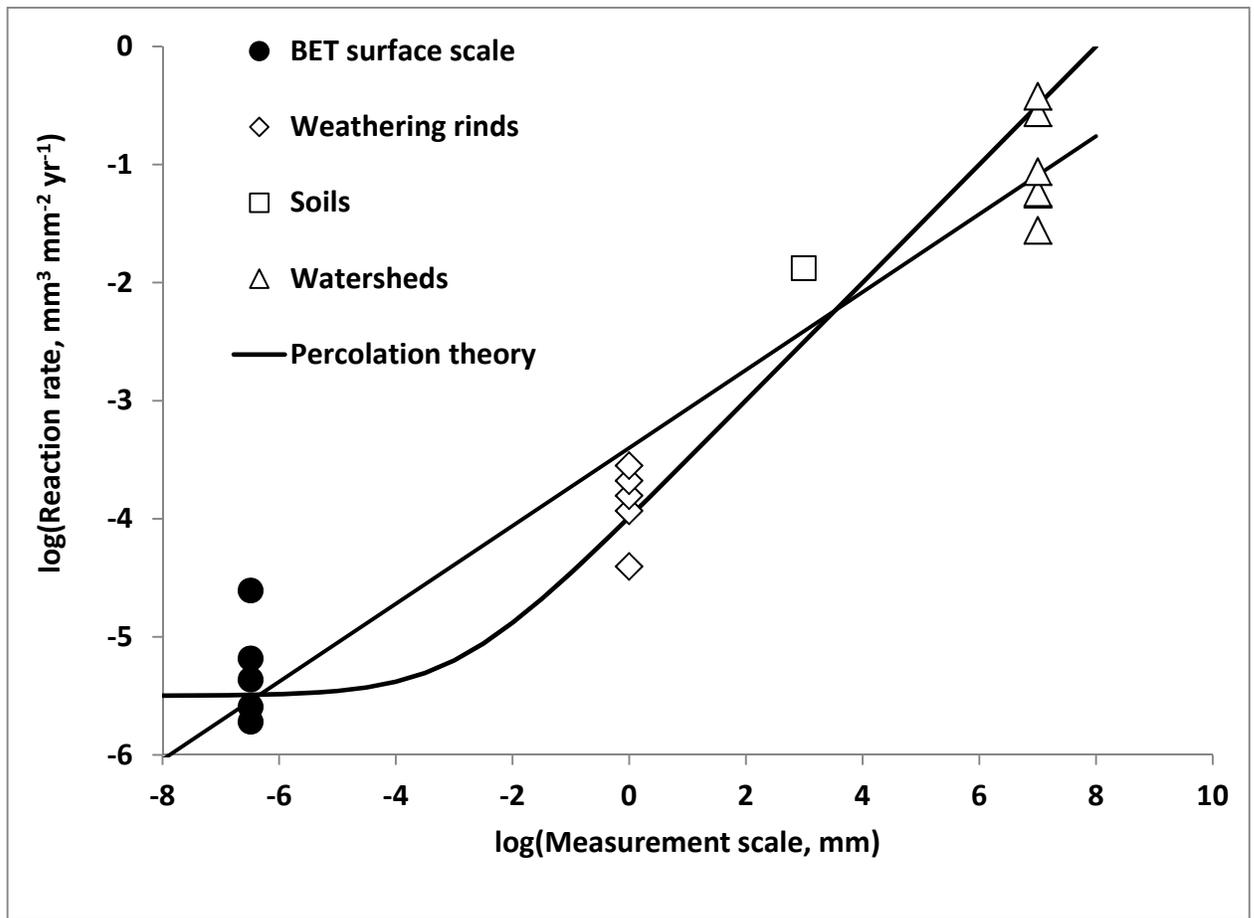

Fig. 4

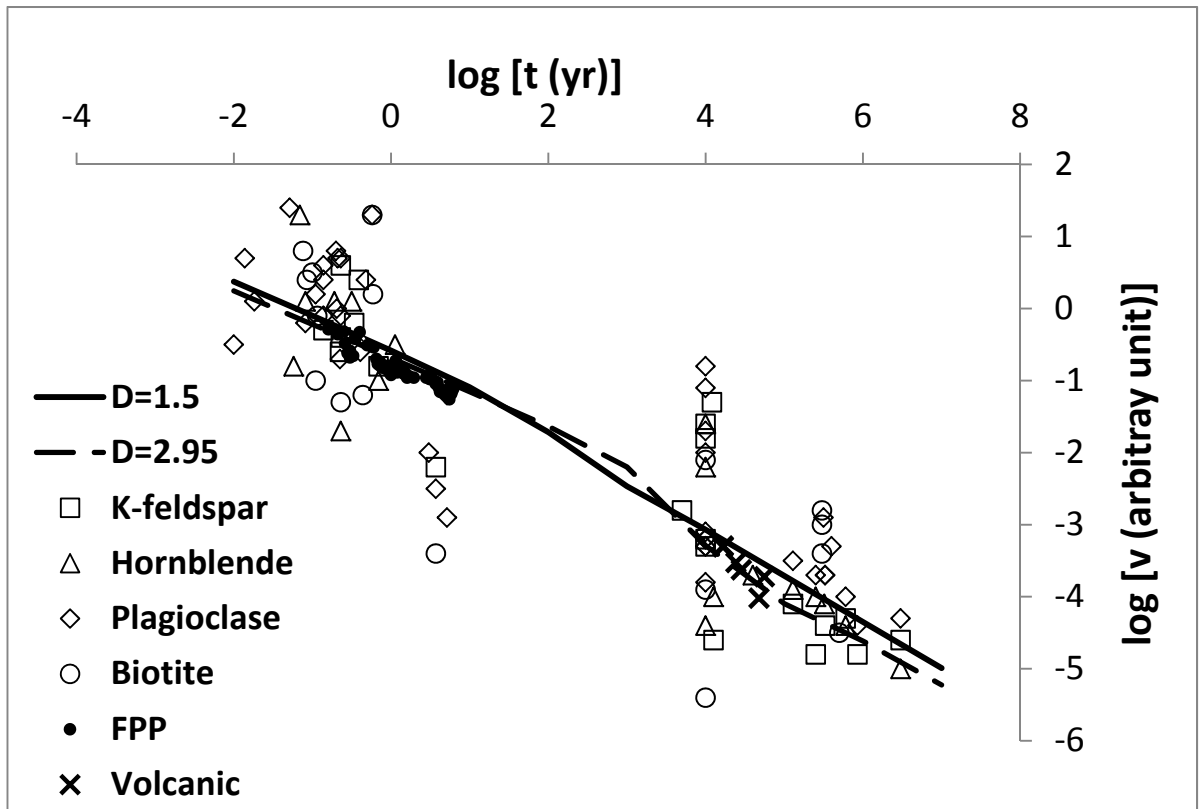
Fig. 5

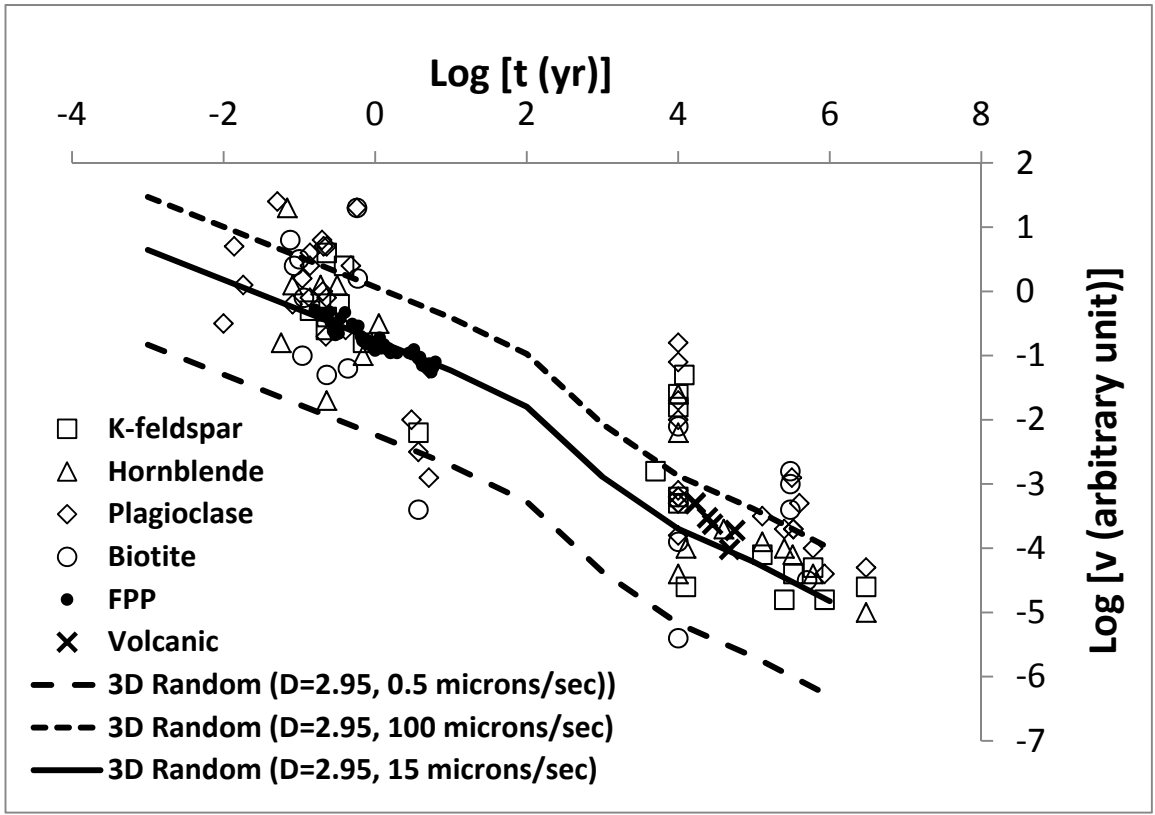

Fig. 6

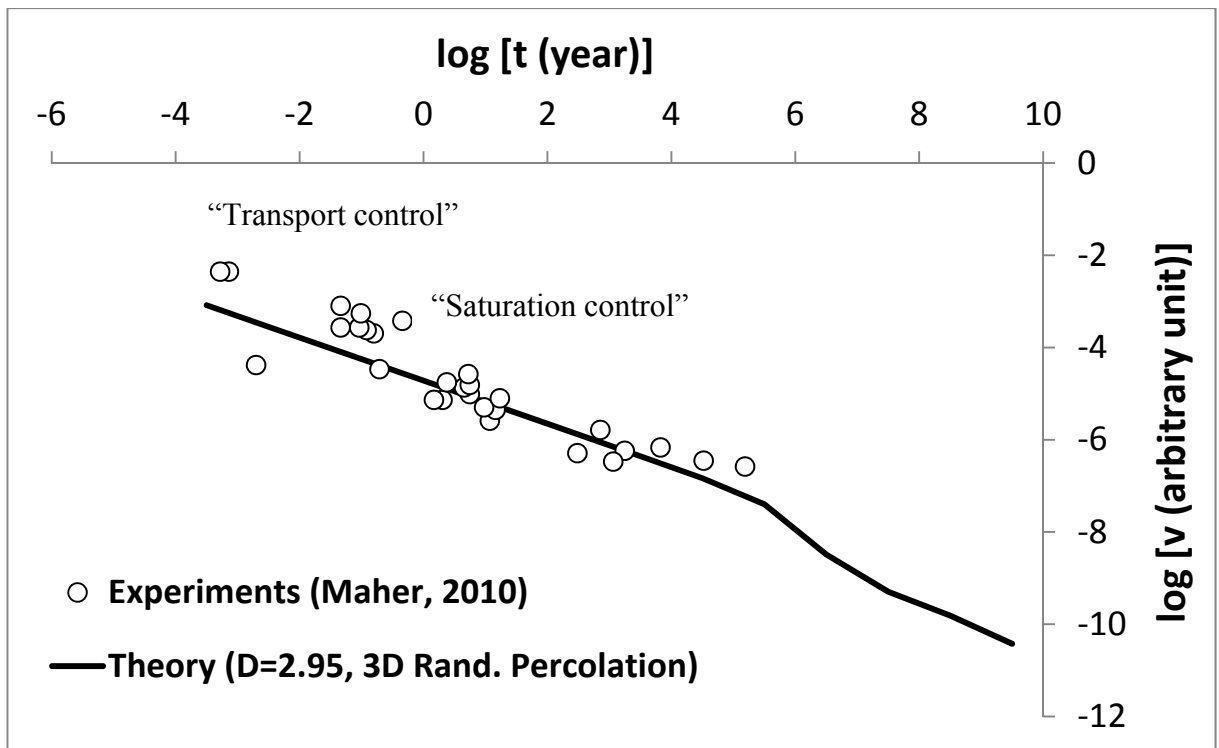

Fig. 7

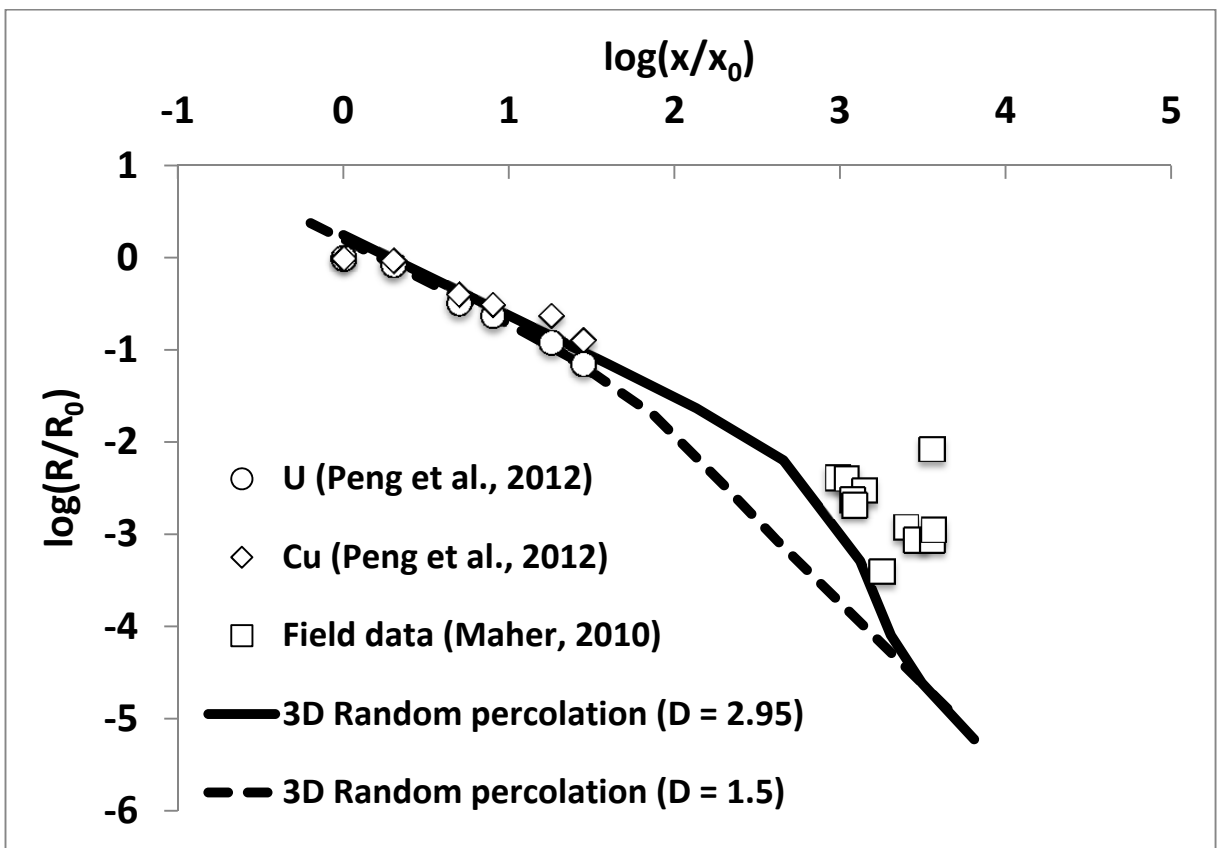

Fig. 8

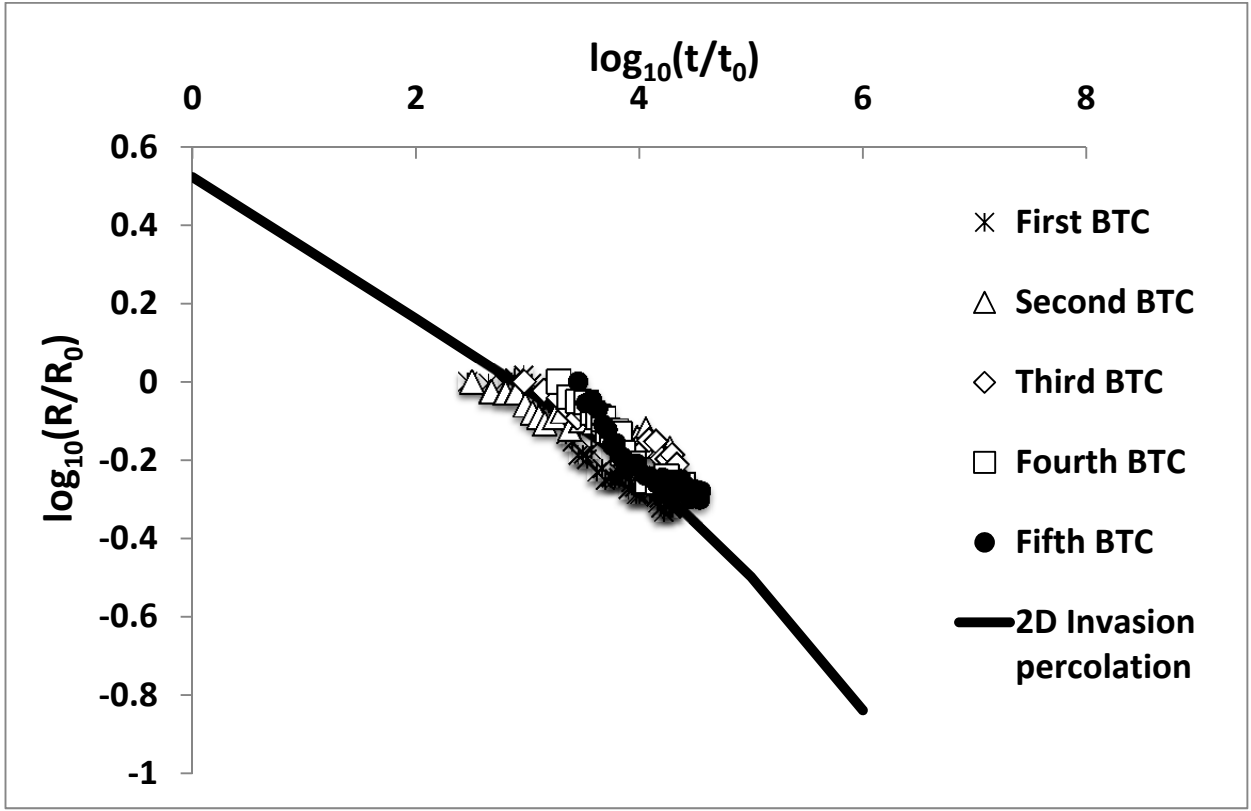

Fig. 9

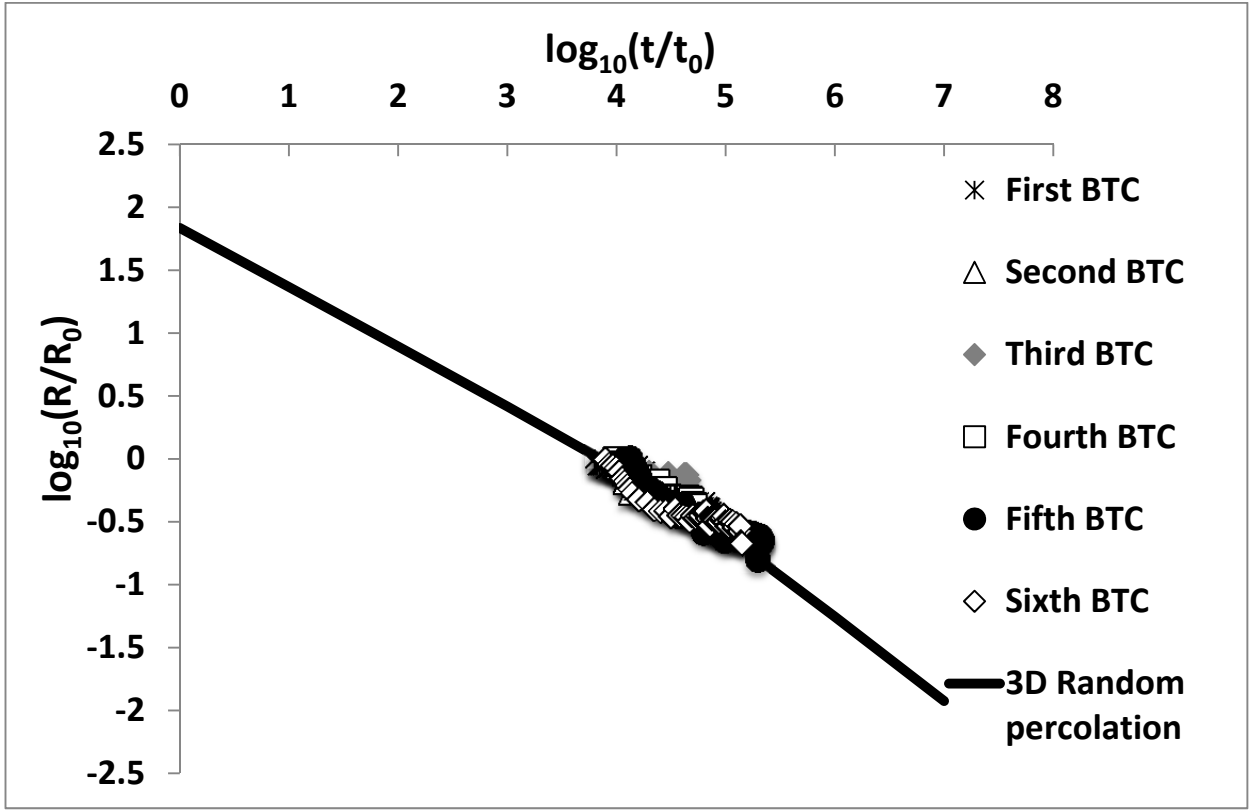

Fig. 10

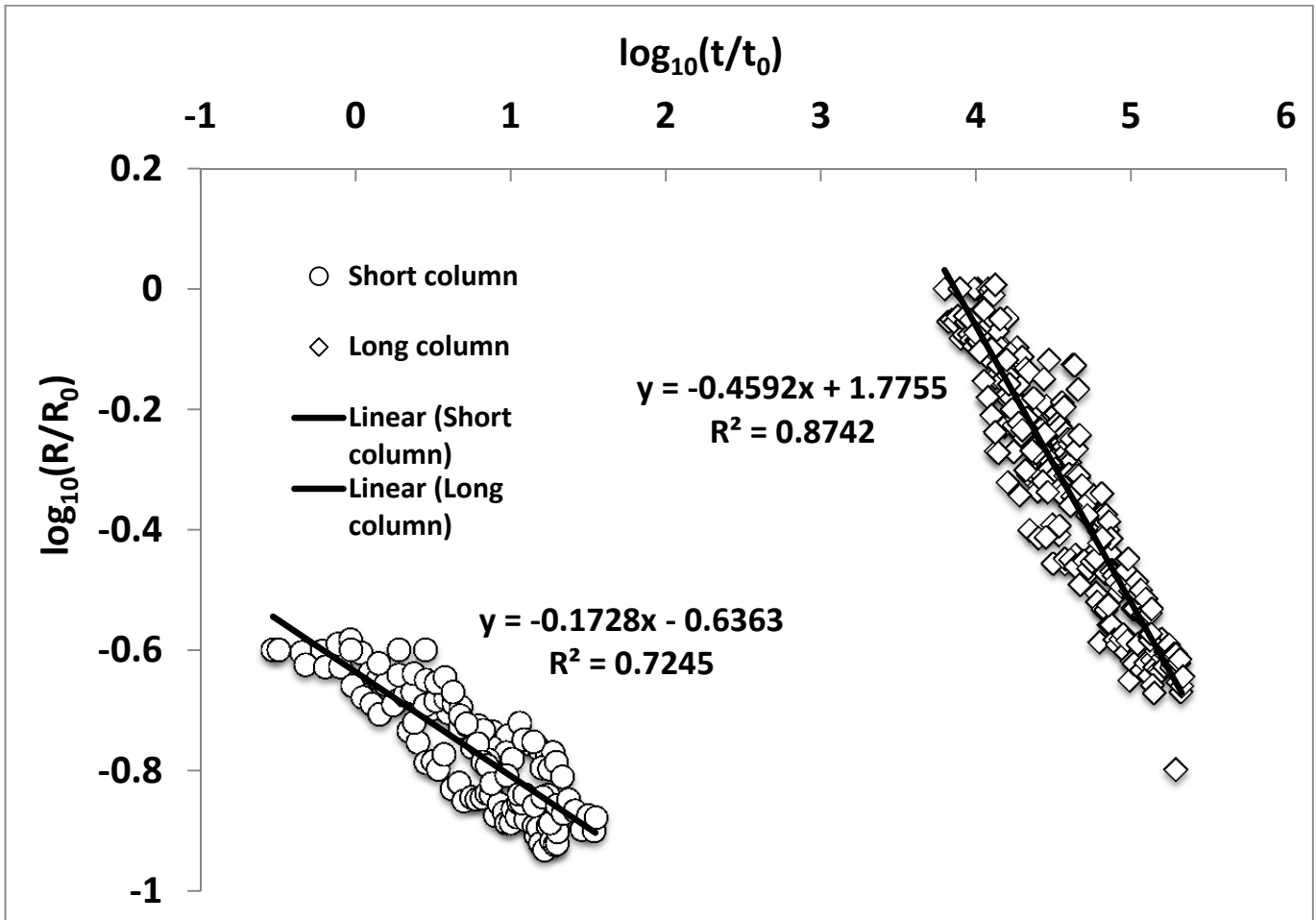
Fig. 11

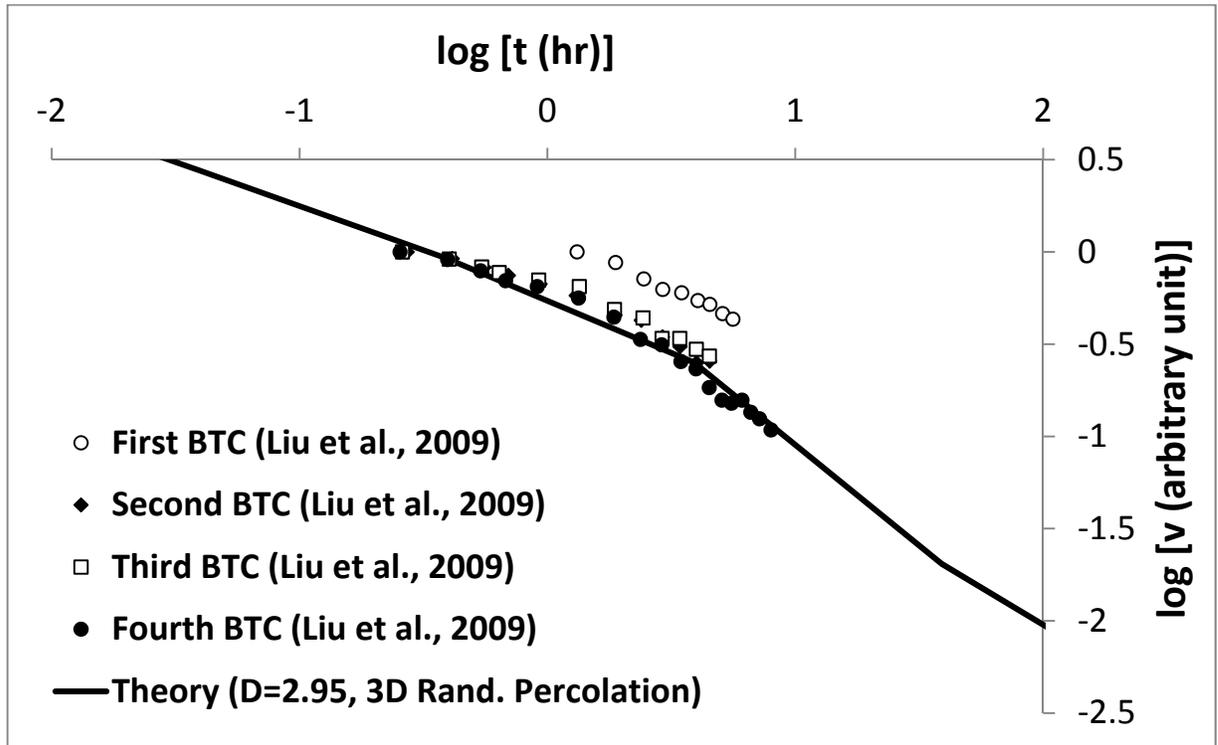

Fig. 12

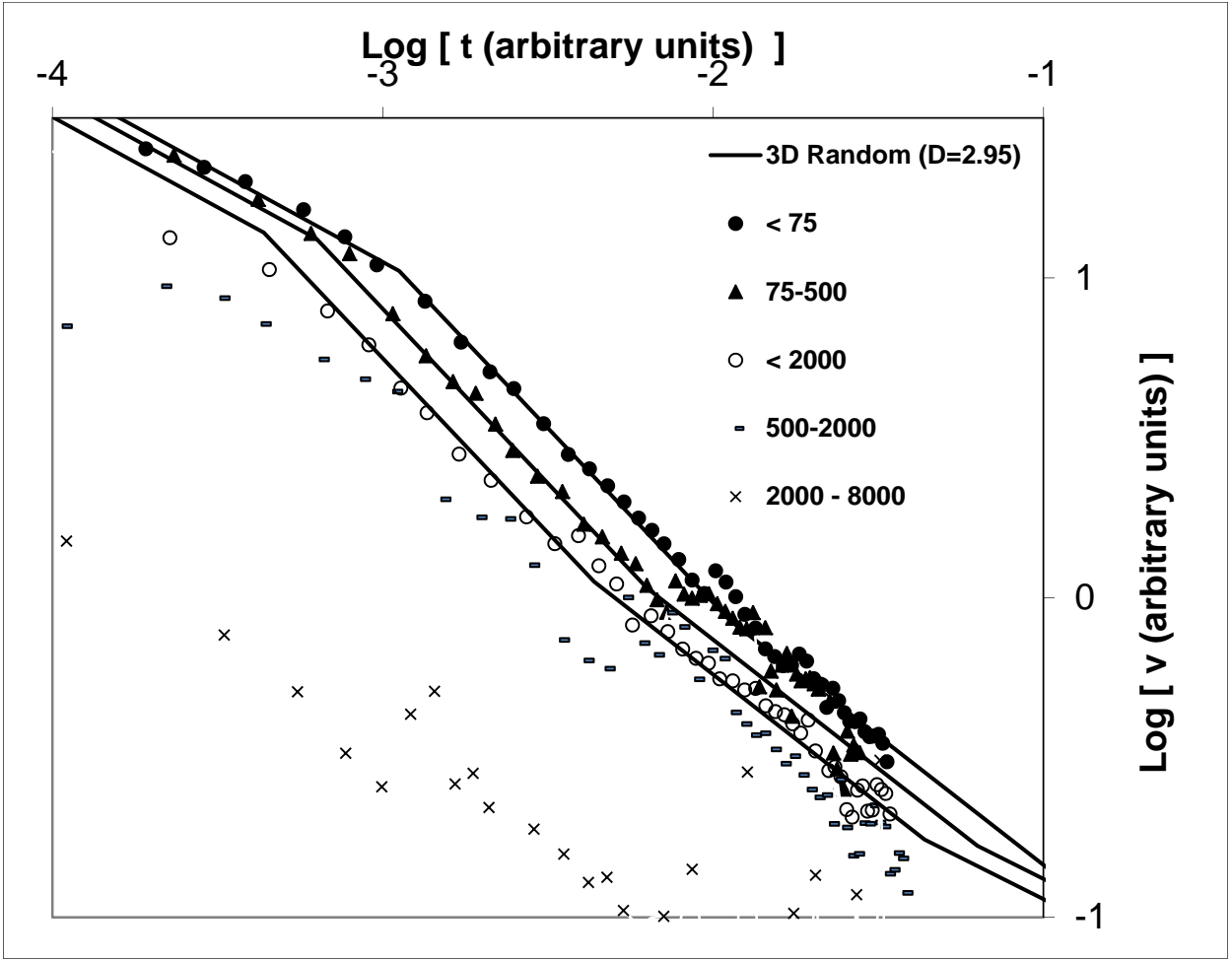

Fig. 13

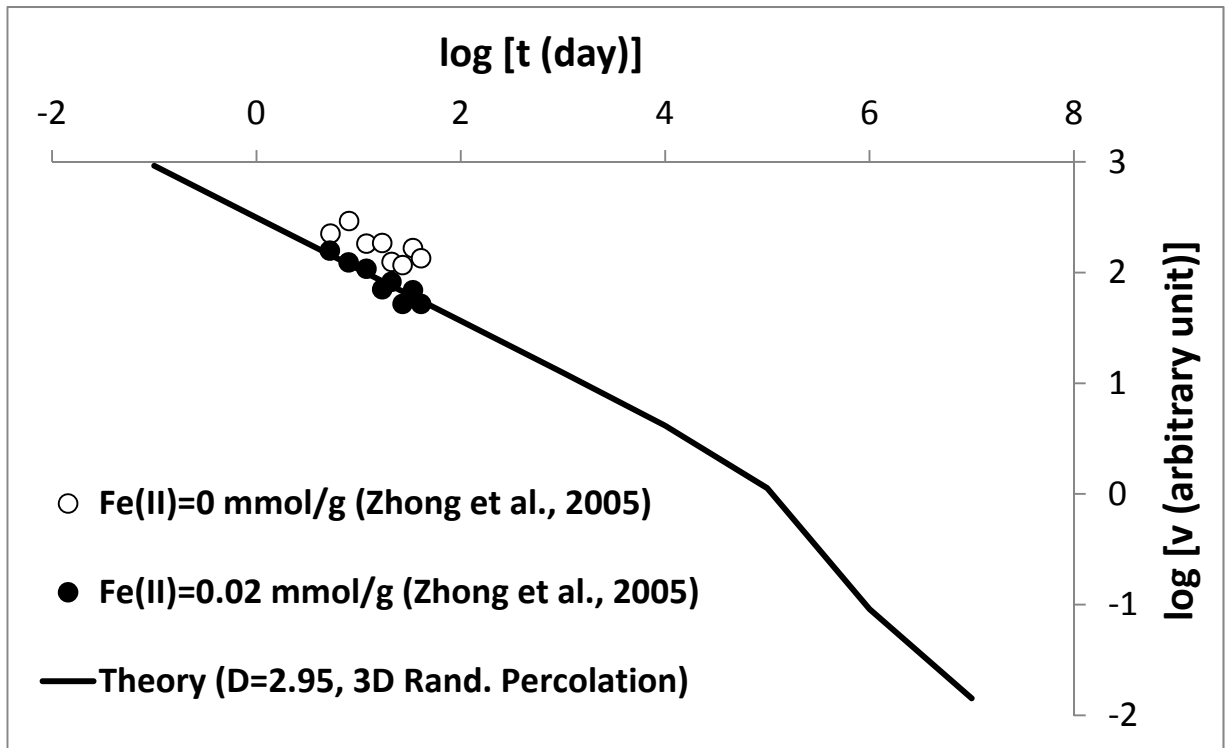

Fig. 14